\documentclass[twocolumn]{aastex631}

\usepackage{amsmath}
\usepackage[encapsulated]{CJK}  % enable CJK characters
\shorttitle{Mid-infrared and CO emission from galaxies}
\shortauthors{Leroy, Bolatto, Sandstrom et al.}

\graphicspath{{./}{figures/}}

\newcommand{\OSU}{\affil{Department of Astronomy, The Ohio State University, 140 West 18th Avenue, Columbus, Ohio 43210, USA}}

\newcommand{\Alberta}{\affil{Department of Physics, University of Alberta, Edmonton, AB T6G 2E1, Canada}}

\newcommand{\ANU}{\affil{Research School of Astronomy and Astrophysics, Australian National University, Canberra, ACT 2611, Australia}}

\newcommand{\ASIAA}{\affil{Institute of Astronomy and Astrophysics, Academia Sinica, No.1, Sec. 4, Roosevelt Rd, Taipei 10617, Taiwan}}

\newcommand{\CCAPP}{\affil{Center for Cosmology and Astroparticle Physics, 191 West Woodruff Avenue, Columbus, OH 43210, USA}}

\newcommand{\CfA}{\affil{Harvard-Smithsonian Center for Astrophysics, 60 Garden Street, Cambridge, MA 02138, USA}}

\newcommand{\CNRS}{\affil{CNRS, IRAP, 9 Av. du Colonel Roche, BP 44346, F-31028 Toulouse cedex 4, France}}

\newcommand{\COOL}{\affil{Cosmic Origins Of Life (COOL) Research DAO, coolresearch.io}}

\newcommand{\ESO}{\affil{European Southern Observatory, Karl-Schwarzschild Stra{\ss}e 2, D-85748 Garching bei M\"{u}nchen, Germany}}

\newcommand{\GEMINI}{\affil{Gemini Observatory/NSF’s NOIRLab, 950 N. Cherry Avenue, Tucson, AZ, 85719, USA}}

\newcommand{\Heidelberg}{\affil{Astronomisches Rechen-Institut, Zentrum f\"{u}r Astronomie der Universit\"{a}t Heidelberg, M\"{o}nchhofstra\ss e 12-14, D-69120 Heidelberg, Germany}}

\newcommand{\ITA}{\affil{Zentrum f\"{u}r Astronomie der Universit\"{a}t Heidelberg, Institut f\"{u}r Theoretische Astrophysik, Albert-Ueberle-Str. 2, D-69120 Heidelberg}}

\newcommand{\IWR}{\affil{Universit\"{a}t Heidelberg, Interdisziplin\"{a}res Zentrum f\"{u}r Wissenschaftliches Rechnen, Im Neuenheimer Feld 205, D-69120 Heidelberg, Germany}}

\newcommand{\JHU}{\affil{Department of Physics and Astronomy, The Johns Hopkins University, Baltimore, MD 21218, USA}}

\newcommand{\ljmu}{\affil{Astrophysics Research Institute, Liverpool John Moores University, 146 Brownlow Hill, Liverpool L3 5RF, UK}}

\newcommand{\Maryland}{\affil{Department of Astronomy and Joint Space-Science Institute, University of Maryland, College Park, MD 20742, USA}}

\newcommand{\MPE}{\affil{Max-Planck-Institut f\"{u}r extraterrestrische Physik, Giessenbachstra{\ss}e 1, D-85748 Garching, Germany}}

\newcommand{\MPIA}{\affil{Max-Planck-Institut f\"{u}r Astronomie, K\"{o}nigstuhl 17, D-69117, Heidelberg, Germany}}

\newcommand{\OAN}{\affil{Observatorio Astron\'{o}mico Nacional (IGN), C/Alfonso XII, 3, E-28014 Madrid, Spain}}

\newcommand{\Oxford}{\affil{Sub-department of Astrophysics, Department of Physics, University of Oxford, Keble Road, Oxford OX1 3RH, UK}}

\newcommand{\UToledo}{\affil{Ritter Astrophysical Center, University of Toledo, 2801 W. Bancroft St., Toledo, OH, 43606}}

\newcommand{\UBonn}{\affil{Argelander-Institut f\"ur Astronomie, Universit\"at Bonn, Auf dem H\"ugel 71, 53121 Bonn, Germany}}

\newcommand{\UCSD}{\affil{Center for Astrophysics and Space Sciences, Department of Physics,  University of California,\\ San Diego, 9500 Gilman Drive, La Jolla, CA 92093, USA}}

\newcommand{\UGent}{\affil{Sterrenkundig Observatorium, Universiteit Gent, Krijgslaan 281 S9, B-9000 Gent, Belgium}}

\newcommand{\UVic}{\affil{Department of Physics \& Astronomy, University of Victoria, Finnerty Road, Victoria, British Columbia, V8P 1A1, Canada}}

\newcommand{\STScIESA}{\affil{AURA for the European Space Agency (ESA), Space Telescope Science Institute, 3700 San Martin Drive, Baltimore, MD 21218, USA}}

\newcommand{\McMaster}{\affil{Department of Physics and Astronomy, McMaster University, Hamilton, ON L8S 4M1, Canada}}

\newcommand{\UA}{\affil{Centro de Astronomía (CITEVA), Universidad de Antofagasta, Avenida Angamos 601, Antofagasta, Chile}}

\newcommand{\MPIFR}{\affil{Max-Planck-Institut f\"ur Radioastronomie, Auf dem H\"ugel 69, 53121 Bonn, Germany}}

\newcommand{\Arizona}{\affil{Steward Observatory, University of Arizona, Tucson, AZ 85721, USA}}

\newcommand{\Indiana}{\affil{Department of Astronomy, Indiana University, Bloomington, IN 47405, USA}}

\newcommand{\UWO}{\affil{Department of Physics \& Astronomy, University of Western Ontario, London, ON N6A 3K7, Canada}}

\newcommand{\CITA}{\affiliation{Canadian Institute for Theoretical Astrophysics (CITA), University of Toronto, 60 St George Street, Toronto, ON M5S 3H8, Canada}}

\newcommand{\NAOJ}{\affiliation{National Astronomical Observatory of Japan, 2-21-1 Osawa, Mitaka, Tokyo, 181-8588, Japan}}

\newcommand{\CRAL}{\affiliation{Univ Lyon, Univ Lyon1, ENS de Lyon, CNRS, Centre de Recherche Astrophysique de Lyon UMR5574, F-69230 Saint-Genis-Laval France}}

\begin{document}

% Submitted on October 22, 2022

\title{PHANGS--JWST First Results: A Global and Moderately Resolved View of Mid-Infrared and CO Line Emission from Galaxies at the Start of the JWST Era }

\begin{abstract}
We explore the relationship between mid-infrared (mid-IR) and CO rotational line emission from massive star-forming galaxies, which is one of the tightest scalings in the local universe. We assemble a large set of unresolved and moderately ($\sim 1$~kpc) spatially resolved measurements of CO~(1-0) and CO~(2-1) intensity, $I_{\rm CO}$, and mid-IR intensity, $I_{\rm MIR}$, at 8, 12, 22, and 24$\mu$m. The $I_{\rm CO}$ vs.\ $I_{\rm MIR}$ relationship is reasonably described by a power law with slopes $0.7{-}1.2$  and normalization $I_{\rm CO} \sim 1$~K~km~s$^{-1}$ at $I_{\rm MIR} \sim 1$~MJy~sr$^{-1}$. Both the slopes and intercepts vary systematically with choice of line and band. The comparison between the relations measured for CO~(1-0) and CO~(2-1) allow us to infer that $R_{21} \propto I_{\rm MIR}^{0.2}$, in good agreement with other work. The $8\mu$m and $12\mu$m bands, with strong PAH features, show steeper CO vs.\ mid-IR slopes than the $22\mu$m and $24\mu$m, consistent with PAH emission arising not just from CO-bright gas but also from atomic or CO-dark gas. The CO-to-mid-IR ratio correlates with global galaxy stellar mass ($M_\star$) and anti-correlates with SFR/$M_\star$. At $\sim 1$~kpc resolution, the first four PHANGS-JWST targets show CO to mid-IR relationships that are quantitatively similar to our larger literature sample, including showing the steep CO-to-mid-IR slopes for the JWST PAH-tracing bands, although we caution that these initial data have a small sample size and span a limited range of intensities.
\end{abstract}

\author[0000-0002-2545-1700]{Adam~K.~Leroy}
\OSU \CCAPP

\author[0000-0002-5480-5686]{Alberto D. Bolatto}
\Maryland

\author[0000-0002-4378-8534]{Karin~Sandstrom}
\UCSD

\author[0000-0002-5204-2259]{Erik~Rosolowsky}
\Alberta

\author[0000-0003-0410-4504]{Ashley.~T.~Barnes}
\affiliation{\UBonn}

\author[0000-0003-0166-9745]{F. Bigiel}
\UBonn

\author[0000-0003-0946-6176]{Médéric Boquien}
\UA

\author[0000-0002-8760-6157]{Jakob~S. den Brok}
\UBonn

\author[0000-0001-5301-1326]{Yixian Cao}
\MPE

\author[0000-0002-5235-5589]{J\'er\'emy~Chastenet}
\UGent

\author[0000-0002-5635-5180]{M\'elanie~Chevance}
\ITA
\COOL

\author[0000-0003-2551-7148]{I-Da Chiang \begin{CJK*}{UTF8}{bkai}(江宜達)\end{CJK*}}
\ASIAA

\author[0000-0001-8241-7704]{Ryan Chown}
\UWO

\author[0000-0001-6498-2945]{Dario Colombo}
\MPIFR

\author[0000-0002-1768-1899]{Sara L. Ellison}
\UVic

\author[0000-0002-6155-7166]{Eric Emsellem}
\ESO
\CRAL

\author[0000-0002-3247-5321]{Kathryn Grasha}
\ANU

\author[0000-0001-9656-7682]{Jonathan~D.~Henshaw}
\ljmu
\MPIA

\author[0000-0002-9181-1161]{Annie~Hughes}
\CNRS
 
\author[0000-0002-0560-3172]{Ralf S.\ Klessen}
\ITA
\IWR

\author[0000-0001-9605-780X]{Eric W. Koch}
\CfA

\author[0000-0002-0432-6847]{Jaeyeon Kim}
\ITA

\author[0000-0001-6551-3091]{Kathryn Kreckel}
\Heidelberg

\author[0000-0002-8804-0212]{J.~M.~Diederik~Kruijssen}
\COOL

\author[0000-0003-3917-6460]{Kirsten~L.~Larson}
\STScIESA

\author[0000-0003-0946-6176]{Janice C. Lee}
\GEMINI
\Arizona

\author[0000-0003-2508-2586]{Rebecca~C.~Levy}
\altaffiliation{NSF Astronomy and Astrophysics Postdoctoral Fellow}
\Arizona

\author[0000-0001-7218-7407]{Lihwai Lin}
\ASIAA

\author[0000-0001-9773-7479]{Daizhong Liu}
\MPE

\author[0000-0002-6118-4048]{Sharon~E.~Meidt}
\UGent

\author[0000-0003-3061-6546]{Jérôme Pety}
\affiliation{IRAM, 300 rue de la Piscine, 38400 Saint Martin d'H\`eres, France}
\affiliation{LERMA, Observatoire de Paris, PSL Research University, CNRS, Sorbonne Universit\'es, 75014 Paris}

\author[0000-0002-0472-1011]{Miguel~Querejeta}
\OAN

\author[0000-0003-2508-2586]{M\'onica.~Rubio}
\affil{Departamento de Astronom\'ia, Universidad de Chile, Casilla 36-D, Santiago, Chile}

\author[0000-0002-2501-9328]{Toshiki Saito}
\NAOJ

\author[0000-0003-2342-7501]{Samir~Salim}
\Indiana

\author[0000-0002-3933-7677]{Eva~Schinnerer}
\MPIA

\author[0000-0001-6113-6241]{Mattia C. Sormani}
\ITA

\author[0000-0003-0378-4667]{Jiayi~Sun}
\McMaster
\CITA

\author[0000-0002-8528-7340]{David A. Thilker}
\JHU

\author[0000-0003-1242-505X]{Antonio Usero}
\OAN

\author[0000-0002-8765-7915]{Stuart N. Vogel}
\Maryland

\author[0000-0002-7365-5791]{Elizabeth~J.~Watkins}
\Heidelberg

\author[0000-0003-2093-4452]{Cory~M.~Whitcomb}
\UToledo

\author[0000-0002-0012-2142]{Thomas~G.~Williams}
\Oxford
\MPIA

\author[0000-0001-5817-0991]{Christine D. Wilson}
\McMaster

\suppressAffiliations


\received{October 25, 2022}
\accepted{December 12, 2022}
\submitjournal{The Astrophysical Journal Letters}

\section{Introduction} 
\label{sec:intro}

In this paper, we characterize the observed relationship between mid-infrared (mid-IR, here spanning $\lambda \approx 8{-}24\mu$m) dust emission and CO $J=1\rightarrow0$ and $J=2\rightarrow1$ line emission at moderate spatial resolution ($\sim 1$~kpc) in low redshift star-forming galaxies. 

Mid-IR emission emerges from small dust grains that are mostly heated by ultraviolet (UV) photons. These small grains likely include polycyclic aromatic hydrocarbons (PAHs) that produce distinctive emission features at $\lambda = 7.7\mu$m and $\lambda = 11.3\mu$m \citep[e.g.,][]{DRAINE11BOOK}. Physically, this emission arises from a combination of PAH band and continuum emission from stochastically heated small dust grains. The mid-IR emission of the dust grains responds directly to the local radiation field. This is usually dominated by UV radiation from young stars, although older stellar populations contribute in a proportion that depends on the environment \citep[e.g.,][]{Liu2011,LEROY12SFR,BENDO12SFR,Crocker2013} and may be sometimes dominant \citep[e.g.,][]{GROVES12DUST,VIAENE17}. Because of the ability of dust to capture and reprocess UV radiation from young stars and the observed excellent correlation between mid-IR, H$\alpha$, and UV emission in galaxies, the mid-IR has been widely used as a star formation tracer \citep[e.g., see reviews in][]{KENNICUTT12REVIEW,CALZETTI13REVIEW}. Because mid-IR reflects the dust surface density in addition to its heating, and because star formation and cold gas tend to track one another well at large scales, the mid-IR has also been used as a tracer of the gas mass or gas distribution in galaxies \citep[e.g.,][]{GAO19MIRCO,CHOWN21SFGAS,PHANGSALMA21,GAO22MIRCO}. Practically, the sensitivity of \textit{Spitzer} \citep{SPITZER04} at $8$ and $24\mu$m and  the all-sky coverage of the WISE satellite \citep{WISE10} at $12\mu$m and $22\mu$m have rendered mid-IR emission widely accessible, so that a large fraction of the work on star formation in nearby galaxies over the last two decades has leveraged mid-IR emission in some way \citep[e.g.,][]{KENNICUTT12REVIEW,GSWLC16,JANOWIECKI17SFR}.

Meanwhile, low-$J$ rotational line emission from CO remains the standard observational tracer of the cold molecular gas in galaxies \citep{BOLATTO13REVIEW}. Thanks to large time investments and major improvements in instrumentation, CO~(1-0) emission at $\lambda = 2.6$~mm and CO~(2-1) emission at $\lambda = 1.3$~mm have been observed from hundreds of local galaxies over the last two decades \citep[e.g.,][]{BIMASONG03,NROSURVEY07,HERACLES09,AMIGA11,COLDGASS11,XCOLDGASS17,ALLSMOG14,ALLSMOG17,ALMAQUESTSURVEY20,COMING19,PHANGSALMA21,VERTICO21,MASCOTSURVEY22}, often with the goal of better understanding the relationship between molecular gas and star formation. Due to the all-sky coverage by WISE and extensive mapping programs by \textit{Spitzer}, most of these galaxies have some form of mid-IR observation \citep[e.g.,][]{SINGSSURVEY03,LVLSURVEY09,WISE10,BENDO12SURVEY,UNWISE14,LEROY19Z0MGS}. As a result, for much of the last two decades, our understanding of star formation and molecular gas in galaxies has been inextricably linked to the observed relationship between mid-infrared and CO line emission.

The successful deployment of the JWST promises to keep this topic in the spotlight and push the field forward in terms of physical resolution and sensitivity. In particular, JWST and ALMA, NOEMA, or the SMA can together map mid-IR and CO emission at $< 1''$ resolution, sufficient to resolve substructure within individual regions in the nearest galaxies and to break more distant galaxies apart into individual regions.

To frame these measurements and inform how to best use JWST and ALMA together, here we take stock of what we currently know about the relationship between mid-IR and CO emission from galaxies. To our knowledge, no one has consistently combined the large available dataset of mapping and integrated surveys of galaxies in CO, targeted \textit{Spitzer} mid-IR mapping, and all-sky mid-IR measurements from WISE into a single coherent analysis. In fact, there has been relatively little direct observational analysis of the mid-IR--CO relation \citep[though][are important exceptions]{CHOWN21SFGAS,GAO22MIRCO,WHITCOMB22MIRCO}, with most studies focusing instead on physical quantities often derived after combination with other bands.

In this paper we compile a large collection of integrated and modest resolution CO line and mid-IR measurements to evaluate how CO scales with mid-IR emission. Specifically, we measure how the mean CO~(2-1) and CO~(1-0) intensities in a galaxy scale with the mean $12\mu$m and $22\mu$m surface brightness inferred from WISE data (\S \ref{sec:scaling}), and we also measure the scaling between CO intensity and mid-IR emission at $8\mu$m, $12\mu$m, $22\mu$m, and $24\mu$m using moderately resolved observations ($\theta = 17\arcsec$ corresponding to a median of $1.3$~kpc and set by the resolution of the mid-IR data).

We approach these measurements from an empirical perspective. Our goal is to synthesize the observational state of this topic heading into the JWST era. With that in mind, we provide typical ratios, scatter, and fits of CO intensity to IR intensity for each combination of mid-IR band and CO line. We compare results for different lines and bands, and examine how the CO to mid-IR ratio depends on global galaxy properties. We connect our results to the recent literature on dust, gas, and star formation in galaxies, but throughout we maintain a firmly observational perspective: our goal is to distill the current state of matched resolution observations of CO and mid-IR emission.

\section{Data} \label{sec:data}

\begin{deluxetable*}{l|c|c|c}[ht!]
\tabletypesize{\small}
\tablecaption{Summary of data sets \label{tab:sample1}}
\tablewidth{0pt}
\tablehead{
\colhead{Data pair} & 
\colhead{CO Det. \& MIR Det.} &
\colhead{CO Lim. \& MIR Det.} &
\colhead{CO Det. \& MIR Lim.}
}
\startdata
\hline
\multicolumn{4}{c}{Integrated galaxies (Figures \ref{fig:intscaling}, \ref{fig:intscaling_mass} \& \ref{fig:intscaling_ssfr})} \\
\hline
CO~(1-0) \& 12$\mu$m & 823 & 313 & 11 \\
CO~(1-0) \& 22$\mu$m & 771 & 173 & 63 \\
CO~(2-1) \& 12$\mu$m & 354 & 187 & 2 \\
CO~(2-1) \& 22$\mu$m & 347 & 133 & 9 \\
\hline
\multicolumn{4}{c}{Rings in radial profiles (Figures \ref{fig:resscalingwise} \& \ref{fig:resscalingspitzer})} \\
\hline
CO~(1-0) \& 12$\mu$m & 980 & 333 & 0 \\
CO~(1-0) \& 22$\mu$m & 980 & 324 & 0 \\
CO~(1-0) \& 8$\mu$m & 357 & 142 & 0 \\
CO~(1-0) \& 24$\mu$m & 456 & 159 & 0 \\
CO~(2-1) \& 12$\mu$m & 1123 & 218 & 0 \\
CO~(2-1) \& 22$\mu$m & 1122 & 158 & 1 \\
CO~(2-1) \& 8$\mu$m & 551 & 177 & 1 \\
CO~(2-1) \& 24$\mu$m & 739 & 230 & 0 \\
\hline
\multicolumn{4}{c}{Individual $17''$ regions (Figures \ref{fig:pixscalingwise} \& \ref{fig:pixscalingspitzer})} \\
\hline
CO~(1-0) \& 12$\mu$m & 4540 & 26459 & 3 \\
CO~(1-0) \& 22$\mu$m & 4503 & 17858 & 40 \\
CO~(1-0) \& 8$\mu$m & 1135 & 6864 & 0 \\
CO~(1-0) \& 24$\mu$m & 1582 & 8937 & 0 \\
CO~(2-1) \& 12$\mu$m & 13473 & 17444 & 95 \\
CO~(2-1) \& 22$\mu$m & 11989 & 7607 & 1579 \\
CO~(2-1) \& 8$\mu$m & 3647 & 8818 & 26 \\
CO~(2-1) \& 24$\mu$m & 6809 & 11862 & 73 \\
\enddata
\tablecomments{``Detections'' and ``limits'' assigned using a $S/N=5$ threshold. Repeat measurements from different surveys are allowed for integrated galaxies but not resolved measurements. Resolved measurements oversample the beam by a factor of $2$ for radial profiles and $4$ (in area) for individual regions. Our $17''$ resolution corresponds to a median $1.3$~kpc ($16{-}84\%$ range $0.5{-}1.9$~kpc) for the rings in radial profiles and median $1.2$~kpc ($16{-}84\%$ range $0.3{-}1.8$~kpc) for individual regions. See \S \ref{sec:data} for more details and references to individual survey data.}
\end{deluxetable*}

\begin{figure*}
\centering
\includegraphics[width=0.8\textwidth]{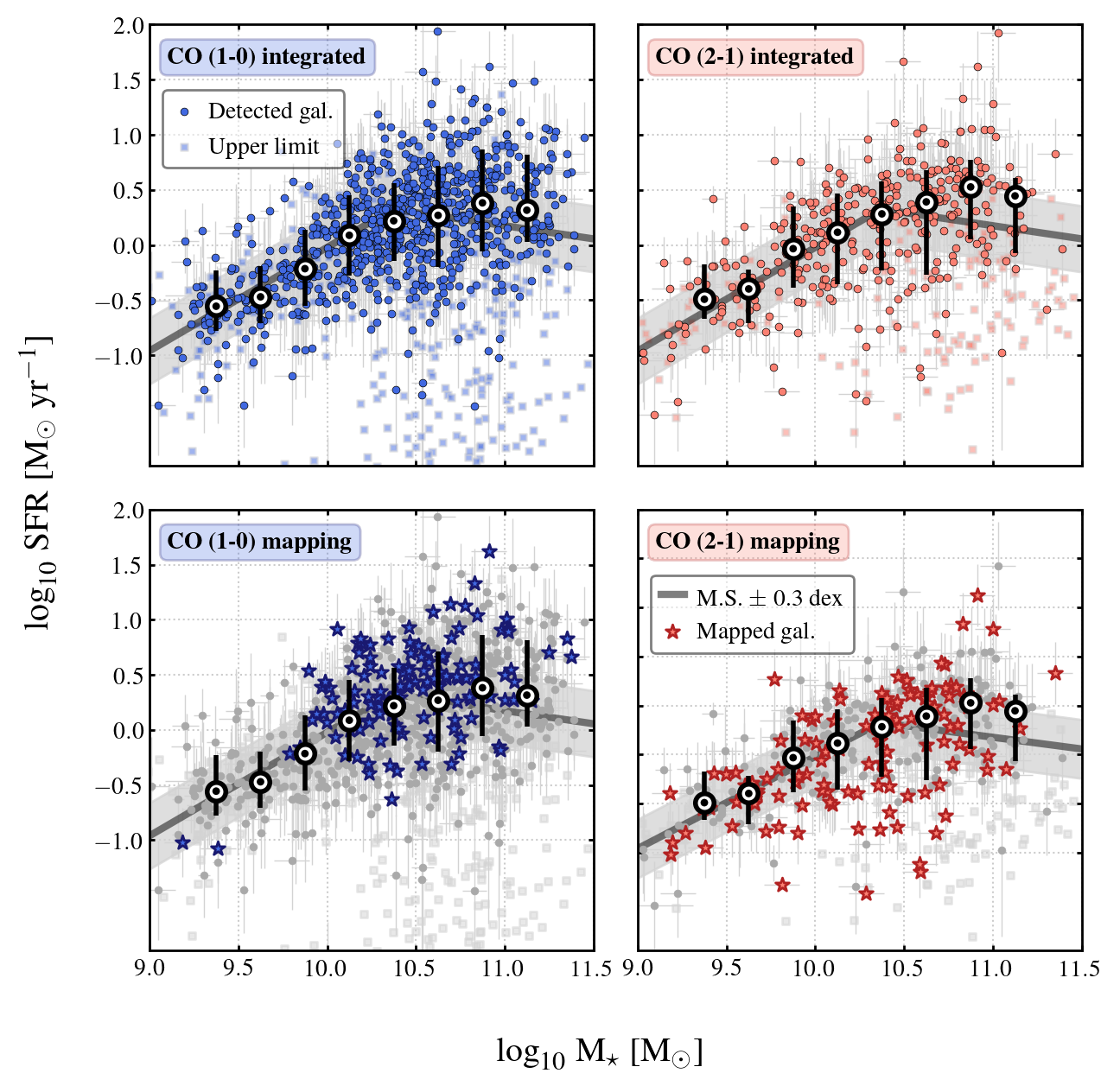}
\caption{\textit{Sample of galaxies studied in star formation rate vs. stellar mass space.} Each point corresponds to one of the galaxies that we study, with solid symbols indicating galaxies detected in both WISE3 and the relevant CO line for that panel. The top row shows the properties of galaxies with integrated measurements, while the bottom row highlights integrated measurements for galaxies with CO mapping. In each panel, we indicate the ``main sequence'' relation between SFR and $M_\star$ from \citet{LEJA22MS} with a $\pm 0.4$~dex range. The black-and-white circles show the median and the corresponding error bars show the 16{-}84\% range of SFR at fixed $M_\star$ for all galaxies detected in WISE3 and the CO line shown in that panel. Overall the integrated CO~(1-0) and CO~(2-1) sample and the mapping CO~(2-1) data cover the star forming main sequence, and we expect that they represent the overall $z=0$ massive, star-forming galaxies well. The CO~(1-0) mapping coverage is somewhat more restricted and slightly biased, a topic that we discuss in detail in \S \ref{sec:intprops}. Low mass dwarf galaxies and low SFR/$M_\star$ early type galaxies are not well-covered by our sample. See Tables \ref{tab:sample1} and \ref{tab:sample2} for more information.
\label{fig:sample}
}
\end{figure*}

In our analysis, we use deprojected (face-on) estimates of the line-integrated CO intensity and the mid-IR intensities at $8\mu$m, $12\mu$m, $22\mu$m, and $24\mu$m. When not referencing a specific line or band, we refer to the CO intensity as $I_{\rm CO}$ and mid-IR intensity as $I_{\rm MIR}$. We derive these numbers using a large set of previously published galaxy-integrated (\S \ref{sec:intdata}) and moderately resolved (\S \ref{sec:resdata}) maps of CO and mid-IR emission from nearby galaxies (c.f., Tables \ref{tab:sample1} and \ref{tab:sample2} and Figure \ref{fig:sample}).
 
We use units of K~km~s$^{-1}$ to describe the CO intensity and record separate results for the CO $J=1\rightarrow0$ transition at $\nu = 115.27$~GHz and CO~$J=2\rightarrow1$ at $\nu = 230.54$~GHz. For reference we note that for a standard ``Galactic'' CO-to-H$_2$ conversion factor of $\alpha_{\rm CO} = 4.35$ M$_\odot$~pc$^{-2}$ (K~km~s$^{-1}$)$^{-1}$ \citep{BOLATTO13REVIEW} and a typical CO~(2-1) to CO~(1-0) ratio of $R_{\rm 21} = 0.65$ \citep{DENBROK21,YAJIMA21,LEROY22}, $I_{\rm CO}$ maps to the molecular gas mass surface density, including helium, via

\begin{equation}
\label{eq:mmolco}
\Sigma_{\rm mol}\approx \begin{cases}
4.35~{\rm M}_\odot~{\rm pc}^{-2} ~ \left( \frac{I_{\rm CO (1-0)}}{{\rm K~km~s^{-1}}} \right) \\
6.7~{\rm M}_\odot~{\rm pc}^{-2} ~ \left( \frac{0.65}{R_{21}} \right) ~ \left( \frac{I_{\rm CO (2-1)}}{{\rm K~km~s^{-1}}} \right)~.
\end{cases}
\end{equation}

We compare CO intensities to observed mid-IR intensities, which we record in units of MJy~sr$^{-1}$. These measurements are available for all galaxies in our sample at $12\mu$m and 22$\mu$m from the WISE satellite. The $8\mu$m and $24\mu$m measurements are available for many nearby galaxies from \textit{Spitzer} and we include these bands in the resolved analysis.

For reference, we note typical conversions between mid-IR brightness at $12\mu$m and $22\mu$m and star formation rate surface density

\begin{equation}
\label{eq:sfrmidir}
\Sigma_{\rm SFR}\approx \begin{cases} 
6.0 \times 10^{-3} {\rm M_\odot~yr^{-1}~kpc^{-2}} \left( \frac{I_{\rm 12\mu m}}{{\rm MJy~sr^{-1}}} \right) \\
3.8 \times 10^{-3} {\rm M_\odot~yr^{-1}~kpc^{-2}} \left( \frac{I_{\rm 22\mu m}}{{\rm MJy~sr^{-1}}} \right)~.
\end{cases}
\end{equation}

\noindent The conversions above are the average ones from \citet{LEROY19Z0MGS} and resemble those in \citet{JARRETT13SFR}, \citet{CATALAN15SFR}, or \citet{JANOWIECKI17SFR}. Details related to the conversion and departures from linearity can be found in \citet{BELFIORE22SFR} and \citet{BOQUIEN21}.

\subsection{Galaxy-integrated data}
\label{sec:intdata}

We compile a large set of galaxy-integrated estimates of CO luminosity, mid-IR luminosity, effective radius, stellar mass, and star formation rate. Given the sky-wide coverage of WISE, the limiting factor is the availability of CO data, so we construct the sample around CO surveys. We compile galaxy-integrated CO~(1-0) luminosities from  ALMaQUEST \citep[][]{ALMAQUESTSURVEY20}, AMIGA \citep[][]{AMIGA11}, CARMA--EDGE \citep[][]{EDGECALIFA17}, MASCOT \citep[][]{MASCOTSURVEY22}, xCOLD~GASS \citep[][]{XCOLDGASS17}, and galaxy-integrated CO~(2-1) luminosities from APEX--EDGE \citep[][]{EDGECALIFA20}, ALLSMOG \citep[][]{ALLSMOG14,ALLSMOG17}, and a survey by \citet[][]{JIANG15CO}. We also use the galaxy-integrated values from the CO~(2-1) and CO~(1-0) mapping samples COMING \citep{COMING19}, HERACLES \citep{HERACLES09} with the additional IRAM 30-m maps described in \citet{LEROY22}, PHANGS--ALMA \citep{PHANGSALMA21}, and the Nobeyama atlas by \citet{NROSURVEY07}. The calculation of these integrated values for mapping surveys are described in \citet{LEROY22} and when necessary includes an aperture correction to convert the mapped luminosity to the full galaxy CO luminosity \citep[e.g., see][]{PHANGSALMA21}. 

We assign mid-IR luminosities at $12\mu$m and $22\mu$m, stellar mass ($M_\star$), and star formation rate (SFR) estimates by cross-matching each galaxy with a CO measurement through either the GSWLC database \citep{GSWLC16,GSWLC18} or the $z0MGS$ atlas \citep{LEROY19Z0MGS}. In the handful of cases where we could not successfully cross-match a CO survey target to either of these mid-IR catalogs, we dropped the target from our analysis.

GSWLC compiled WISE catalog photometry and SED-modeling-based stellar masses and SFR estimates for more distant galaxies, while $z0$MGS constructed images, carried out photometry, and estimated SFR and M$_\star$ from UV, near-IR, and mid-IR for extended local systems. The $z0$MGS recipes to estimate SFR and $M_\star$ were designed to yield values that match the GSWLC values, so all of the integrated values in this paper should be self-consistent and anchored to the results of \citet{GSWLC16} and \citet{GSWLC18}, which use CIGALE \citep{CIGALE19} modeling and the \citet{BRUZUAL03SFR} models.

We also assign an effective radius, $R_e$, to each galaxy. For galaxies at distance $< 50$~Mpc, we prefer near-IR based effective radius measurements by \citet{MUNOZMATEOS15S4G} or \citet{PHANGSALMA21}. When these are not available, we estimate the effective radius from the optical $R_{25}$ in the LEDA database \citep{LEDA03,LEDA14} using a scaling relation in which the ratio $R_e/R_{25}$ depends on stellar mass \citep[derived from the data in][]{MUNOZMATEOS15S4G}. For more distant galaxies, we draw the effective radius from the survey itself. For the SDSS-based surveys these tend to be based on $r$-band, while for CALIFA these are based on stellar mass estimates from the modeling of the integral field unit spectroscopy \citep{SANCHEZ2016}. Taking effective radius measurements from different literature sources adds a small systematic uncertainty that does not have a large impact on our results.

We use these data to estimate the mean $I_{\rm CO}$, $I_{\rm 12\mu m}$, and $I_{\rm 22\mu m}$ within $R_e$ for each target. To do this, we assume that half of the luminosity in CO and mid-IR emission emerges from within the stellar effective radius, $R_e$, and calculate $I_x = 0.5~L_x / \pi R_e^2$ for each band $x$. Such an assumption is well-justified, on average, for star-forming galaxies where the CO, stellar, and star formation distributions tend to track one another quite well \citep[e.g.,][though see especially the latter for some important subtleties]{YOUNG91REVIEW,FCRAOSURVEY95,REGAN06MIRCO,LEROY08SFGAS,HERACLES09,EDGECALIFA17,ALMAQUESTSURVEY20,VERTICO21}. 

For the integrated galaxies, we treat observations with $S/N < 5$ as non-detections and record their $5\sigma$ value as an upper limit. In practice, this affects CO much more than the mid-IR. For this part of the analysis, we also allow repeat measurements in the case that different surveys target the same galaxy.

\subsection{Moderately resolved data}
\label{sec:resdata}

We also assemble a large set of $17\arcsec$ resolution measurements of CO emission and mid-IR emission that resolve nearby galaxies into individual regions. We target $17\arcsec$ because this represents the common resolution achievable by WISE $22\mu$m data ($\theta \approx 15''$\footnote{This represents the nearest ``safe'' Gaussian resolution that can be achieved for WISE4 following \citet{ANIANOKERNELS11}.}) and much of the available single-dish CO mapping ($\theta \approx 15{-}17''$). We smooth to a common Gaussian beam with full width at half maximum (FWHM) of $17\arcsec$, which allows us to combine essentially all recent large CO~(1-0) and CO~(2-1) mapping surveys, \textit{Spitzer} IRAC 8$\mu$m and MIPS 24$\mu$m mapping, and WISE $12\mu$m and $22\mu$m mapping.

We use CO maps from BIMA SONG \citep[CO~(1-0),][]{BIMASONG03}, COMING \citep[CO~(1-0),][]{COMING19}, HERACLES \citep[CO~(2-1),][]{HERACLES09} and the follow up IRAM observations described in \citet{LEROY22}, the Nobeyama survey \citep[CO~(1-0),][]{NROSURVEY07}, PHANGS--ALMA \citep[CO~(2-1),][]{PHANGSALMA21}, and VERTICO \citep[CO~(2-1),][]{VERTICO21}. Wherever possible we use the combined interferometric and total power data: all sources in BIMA SONG, PHANGS--ALMA, and many in VERTICO include total power \citep[see table in][]{VERTICO21}. These surveys sometimes target the same galaxy, and for this resolved part of the analysis we only use a single map per galaxy per transition, with our preferences set by the combination of calibration quality, field of view, and sensitivity. For CO~(1-0) our priority is the COMING survey over the survey by \citet{NROSURVEY07} over BIMA SONG. For CO~(2-1) we prefer PHANGS--ALMA over the IRAM/HERACLES survey over VERTICO.

We combine these CO maps with WISE $12\mu$m and $22\mu$m images from \citet{LEROY19Z0MGS} based on \citet{UNWISE14}. Whenever available, we also include IRAC 8$\mu$m emission from either the LVL \citep[][]{LVLSURVEY09} or SINGS \citep[][]{SINGSSURVEY03} surveys and MIPS $24\mu$m emission from LVL, SINGS, or the data compilation by \citet{BENDO12SURVEY}.

We convolve all data to a common Gaussian resolution of $17\arcsec$ at FWHM, using the kernels by \citet{ANIANOKERNELS11} when necessary to convert from the PSF of the instrument to a Gaussian. At the distances to our targets, this $17\arcsec$ resolution corresponds to a $16{-}84\%$ range of $0.5{-}1.9$~kpc, with a median of $1.3$~kpc, for the sample of azimuthal averages and a $16{-}84\%$ range of $0.3{-}1.8$, with a median of $1.2$~kpc, for the individual regions.

For the CO data, we mostly follow the procedures described in \citet{PHANGSPIPELINE21} and \citet{LEROY22} to convert from data cubes into maps with associated uncertainties. The only important difference is that for this work we create moment maps by integrating the line over a single velocity window for each galaxy tailored to encompass the emission. That is, we create simple ``flat'' moment maps that enable robust averaging. For the mid-IR maps, we estimate the uncertainties in the convolved data from empty regions of the sky near the galaxy. We project all data for each galaxy onto a common astrometric grid with $8.5\arcsec$ pixels, roughly Nyquist sampling the $17\arcsec$ beam. 

We exclude galaxies with inclination $>70^\circ$ from the resolved analysis (no inclination cut is applied in the integrated analysis). All resolved intensities are corrected for the effects of inclination by applying a factor of $1/\cos i$, with the inclinations drawn from \citet{LANG20} when available or otherwise based on isophotal axial ratios used in \citet{MUNOZMATEOS15S4G} or from LEDA \citep{LEDA03}.

We then use the convolved, inclination-corrected data to construct azimuthally averaged intensity profiles for each target. We average intensities in elliptical annuli of width equal to the FWHM beam size and spaced by one half beam width, so that we oversample the beam radially by a factor of $2$. The orientations for these annuli are drawn from the same sources as the inclination. The averages exclude data within $\pm 30^\circ$ of the minor axis to avoid the most severe projection oversampling effects.

We propagate errors from the original maps and mark rings and pixels with S/N$>5$ as detections (as mentioned below the S/N in CO is the limiting criterion for the overwhelming majority of the data). We retain information on the cases with S/N$<5$ so that we can still use these to average and determine integrated trends but consider these individual measurements unreliable.

\subsection{Sample properties}
\label{sec:sample}

\begin{deluxetable}{l|c|c|c}[ht!]
\tabletypesize{\small}
\tablecaption{Summary of properties of CO-detected galaxies \label{tab:sample2}}
\tablewidth{0pt}
\tablehead{
\colhead{Quantity} & 
\colhead{$\log_{10} M_\star$} &
\colhead{$\log_{10}$~SFR} &
\colhead{$\Delta $MS}
}
\startdata
\multicolumn{4}{c}{Integrated CO~(1-0)} \\
\hline
Minimum & 8.4 & -3.2 & -3.3 \\
5\% & 9.5 & -0.7 & -0.6 \\
16\% & 9.9 & -0.4 & -0.4 \\
50\% & 10.4 & 0.1 & 0.0 \\
84\% & 10.9 & 0.6 & 0.4 \\
95\% & 11.2 & 1.0 & 0.8 \\
Maximum & 11.6 & 1.9 & 1.7 \\
\hline
\multicolumn{4}{c}{Integrated CO~(2-1)} \\
\hline
Minimum & 8.7 & -1.5 & -1.7 \\
5\% & 9.4 & -0.8 & -0.7 \\
16\% & 9.7 & -0.4 & -0.4 \\
50\% & 10.3 & 0.2 & 0.1 \\
84\% & 10.8 & 0.6 & 0.4 \\
95\% & 11.0 & 0.9 & 0.7 \\
Maximum & 11.4 & 1.9 & 1.8 \\
\hline
\multicolumn{4}{c}{Mapping CO~(1-0)} \\
\hline
Minimum & 9.2 & -1.1 & -0.9 \\
5\% & 10.0 & -0.3 & -0.5 \\
16\% & 10.1 & 0.0 & -0.2 \\
50\% & 10.5 & 0.4 & 0.2 \\
84\% & 10.8 & 0.8 & 0.6 \\
95\% & 11.0 & 1.0 & 0.8 \\
Maximum & 11.4 & 1.6 & 1.4 \\
\hline
\multicolumn{4}{c}{Mapping CO~(2-1)} \\
\hline
Minimum & 8.7 & -1.4 & -1.7 \\
5\% & 9.4 & -0.9 & -1.0 \\
16\% & 9.7 & -0.5 & -0.5 \\
50\% & 10.4 & 0.0 & 0.0 \\
84\% & 10.8 & 0.6 & 0.4 \\
95\% & 11.0 & 0.8 & 0.6 \\
Maximum & 11.4 & 1.6 & 1.4 \\
\enddata
\tablecomments{Statistics of galaxies with a CO detection and a WISE3 $12\mu$m detection (see Table \ref{tab:sample1} and Figure \ref{fig:sample}. We quote stellar mass ($\log_{10} M_\star$ in $M_\odot$), star formation rate ($\log_{10}$ SFR in $M_\odot$~yr$^{-1}$), and offset of the SFR from that predicted given the stellar mass and the star forming main sequence relation of \citet{LEJA22MS} ($\Delta$ MS in dex).}
\end{deluxetable}

We have compiled a large fraction of the local universe CO~(1-0) and CO~(2-1) observations of massive, star-forming galaxies. A detailed analysis of the effects of combining these different surveys is beyond the scope of this paper, but Tables \ref{tab:sample1} and \ref{tab:sample2} and Figure \ref{fig:sample} provide an overview of the resulting data set. The figures and Table \ref{tab:sample2} show the stellar mass and SFR for CO detected galaxies and compare these to the ``main sequence'' relating SFR to $M_\star$ for low redshift star-forming galaxies \citep[here we use the broken power law form from][]{LEJA22MS}. 

Our integrated-galaxy CO detections span from about $\log_{10} M_\star \sim 9.5{-}11$ and mostly lie near the star-forming main sequence. Although there are significantly more galaxy-integrated CO~(1-0) measurements than CO~(2-1) measurements, both data sets do a good job of spanning this range. The CO~(2-1) mapping covers the same range and also samples the star-forming main sequence well\footnote{Because the integrated-galaxy values for the CO mapping data come from \citet{LEROY22} these omit VERTICO, which had not yet been released. This should not have a significant impact on our results.}. The CO~(1-0) mapping data cover a narrower range of masses and show some bias towards high SFR at fixed $M_\star$. We discuss this in more detail and correct our results for this effect in \S \ref{sec:intprops}.

We focus on massive, star-forming galaxies in our compilation. Although the data set includes a few early-type and low SFR/$M_\star$ galaxies, many of these are CO non-detections. We do not include CO mapping or galaxy-integrated surveys exclusively focused on early type galaxies \citep[e.g.,][]{COMBES07CO,YOUNG11CO,CROCKER11CO,ALATALO13CO,DAVIS14SFGAS}. Nor do we probe far into the dwarf galaxy regime. For example, we do not include measurements by \citet{SCHRUBA12CO,HUNT15CO} or similar work. Finally, we also do not include dedicated surveys of major mergers or luminous/ultraluminous infrared galaxies \citep[e.g.,][]{LISENFELD19CO,HERRERO-ILLANA20}. Extending this work to such targets will be interesting in future work.

\subsection{Methods}
\label{sec:methods}

Our goal is to characterize the observed relationship between each CO line and each mid-IR band for integrated and moderately resolved galaxies. We aim to measure the strength of the correlation, typical ratio and scatter, and to provide basic functional relationships linking the two.

To do this, we construct a series of variable pairs, separately combining CO~(1-0) and CO~(2-1) intensity with each mid-IR band (see Tables \ref{tab:sample1} and \ref{tab:results}). For integrated galaxies, we focus on WISE $12\mu$m and $22\mu$m, which are available for the whole sample\footnote{Although \textit{Spitzer} covered many galaxies, the overlap of that coverage with the integrated-galaxy CO surveys yields a much smaller sample than the overlap of the integrated-galaxy CO surveys with WISE. However the overlap between the \textit{Spitzer} imaging and the CO mapping surveys is more extensive (see Table \ref{tab:sample1}). Therefore we focus the integrated part of the analysis on WISE and use both WISE and \textit{Spitzer} for the moderately resolved part.}. For these integrated galaxies, we also pair the CO-to-mid-IR ratio with the stellar mass, $M_\star$, and with the specific star formation rate, SFR/$M_\star$, estimated as described in \S \ref{sec:intdata}. For resolved galaxies, we also pair each CO line with the available \textit{Spitzer} $8\mu$m and $24\mu$m data. 

We treat the mid-IR emission as the independent (i.e., $x$-axis) variable in the analysis. Table \ref{tab:sample1} shows that our data often include high S/N mid-IR data with low S/N CO measurements but that the reverse is almost never true. This makes ordering the analysis around the mid-IR measurements convenient. When comparing the CO-to-mid-IR ratio to $M_\star$ or SFR/$M_\star$ we treat the latter as the independent variable, since we have SFR and $M_\star$ estimates for all targets (\S \ref{sec:intdata}).

Treating these variables as independent we bin in $x$, usually mid-IR emission, almost always using 0.25~dex-wide bins and then measure the median and $16{-}84\%$ range for the dependent ($y$-axis) variable, usually CO intensity. These calculations take into account upper limits for integrated galaxies. For resolved measurements we include the low $S/N$ data directly, and rely on the averaging performed in binning to yield a good $S/N$ estimate of the median. In the Figures we show estimates of the $16{-}84\%$ range when the lower bound is not a limit and there are at least $12$ detected measurements contributing to the bin. We show the median estimate when it is not a limit and there are at least $6$ measurements contributing. Overall the bins show an excellent match to the ridgeline and extent of the data (e.g., Figure \ref{fig:intscaling}). They offer a simple way to characterize general trends for a large dataset, and a robust way to deal with the frequently modest $S/N$ in the CO mapping data.

The $I_{\rm CO}$ vs.\ $I_{\rm MIR}$ relation appears to be reasonably well described as a power law, at least to first order (\S\ref{sec:results}). We characterize this by fitting a line to the binned data in log-log space. In detail, we use an orthogonal distance minimization to fit

\begin{equation}
\label{eq:coirline}
\log_{10} I_{\rm CO} = m_{\rm CO-IR} \log_{10} I_{\rm IR} + b_{\rm CO-IR}~,
\end{equation}

\noindent where $I_{\rm CO}$ and $I_{\rm MIR}$ are the median CO line intensity in the bin and the mid-IR surface brightness at the bin center for the line and band in question in units of K~km~s$^{-1}$ and MJy~sr$^{-1}$. 

The subscript ``CO-IR'' denotes that we are fitting CO as a function of MIR intensity. This choice is driven by the practical considerations discussed above, but it also makes physical sense if one were attempting to build a model to predict CO emission or molecular gas content \citep[e.g.,][]{GAO19MIRCO,CHOWN21SFGAS,GAO22MIRCO}. However, this sense is opposite to that often used to characterize star formation scaling relations, which posit some variant of $\Sigma_{\rm SFR} \propto \Sigma_{\rm gas}^N$ where gas is the independent variable. Because we will also discuss our results with reference to these star formation scaling relations, we note that, algebraically,

\begin{equation}
\label{eq:irco}
m_{\rm IR-CO} = \frac{1}{m_{\rm CO-IR}}~\textrm{and}~b_{\rm IR-CO} = - \frac{b_{\rm CO-IR}}{m_{\rm CO-IR}}~,
\end{equation}

\noindent and we will reference $m_{\rm IR-CO}$ as a close cognate of the slope of star formation scaling relations, often denoted $N$.

When regressing the CO-to-mid-IR ratio against global properties, we fit

\begin{equation}
\label{eq:propline}
\log_{10} \frac{I_{\rm CO}}{I_{\rm MIR}} = m \left(\log_{10} M_\star - x_0\right) + b~,
\end{equation}

\noindent with $M_\star$ in $M_\odot$ or use the analogous equation substituting SFR/$M_\star$ in units of yr$^{-1}$ for $M_\star$. Here $x_0$ is an offset that allows us to quote the intercept near the center of the data range in order to decrease covariance between $m$ and $b$. We use $x_0 = 10$ when fitting $M_\star$ and $x_0 = -10$ when fitting SFR/$M_\star$.

We note a few points. First, our choice to fit the binned data reflects the aim to obtain a general scaling rather than the most precise possible predictor of the existing distribution. In practice, our approach de-emphasizes the large amount of high quality data near the distribution center (see Figure \ref{fig:sample}, Table \ref{tab:sample2}), and weights more the data at low and high mid-IR emission. Indeed, inspection shows that our fits to bins do an excellent job at capturing the ridgeline of the data over the intensity range spanned by our bins, which is our goal. The slopes of most of these relations can still easily change by $\pm 0.1$ by adopting other methodologies, and our choice to bin by mid-IR and then fit the binned data represents the key choice. We emphasize, however, that we apply exactly the same method to all variable pairs. This allows us to robustly compare results across different mid-IR bands and lines, even when the absolute values have methodological uncertainty. We make an approximate estimate of uncertainties associated with our fits \textit{after binning} by adding in quadrature the result of (a) bootstrap resampling the bins and (b) varying the fitting approach among orthogonal distance, ordinary least squares $y$ vs. $x$, and ordinary least squares $x$ vs. $y$. We also report the range of $x$ values over which we make binned measurements. This represents the range over which we expect the fit to apply.

In addition to fitting, for each data pair we also measure the Spearman rank correlation relating $x$ and $y$ for detected ($S/N>5$) data, the median ratio $\log_{10} y/x$, and the scatter in both the ratio and the residuals of individual data about the fit. The main caveat is that the individual data at $17\arcsec$ resolution are quite noisy, especially for CO~(1-0), resulting in a weaker correlation at these scales. Therefore the radial profile data offer our best high $S/N$ view of moderately resolved galaxies.

Finally, we note that relations with internally correlated axes should be interpreted carefully, because one risks ascribing physical meaning to something that may be purely mathematical. Our only formally correlated variable pair is CO/MIR vs.\ SFR/$M_\star$, where the mid-IR has been used as part of the SFR estimates. The relationship is of interest, and so we plot it, but we caution some of the observed anti-correlation may reflect the built-in correlation due to the variable choice. The rest of our axes are formally independent, though obviously closely linked physically.

\section{Results} \label{sec:results}

\begin{deluxetable*}{l|c|c|c|c|c|c|c|c}[ht!]
\tabletypesize{\small}
\tablecaption{Summary of CO vs.\ mid-IR scaling relations \label{tab:results}}
\tablewidth{0pt}
\tablehead{
\colhead{Data pair} & 
\colhead{Rank} &
\colhead{Ratio} &
\colhead{Scatter} &
\colhead{Bin range} &
\colhead{$m$} & 
\colhead{$b$} & 
\colhead{$\sigma_{\rm resid}$} & 
\colhead{Figure} \\
\colhead{($y$ vs.\ $x$)} &
\colhead{corr.} &
\colhead{($\log_{10} y/x$)} &
\colhead{(dex)} &
\colhead{(in $\log_{10} x$)} &
\multicolumn{2}{c}{(Equations \ref{eq:coirline} \& \ref{eq:propline})} &
%\colhead{} &
%\colhead{} &
\colhead{(dex)} & 
\colhead{}
}
\startdata
\hline
\multicolumn{8}{c}{Integrated galaxies} \\
\hline
CO~(1-0) vs.\ 12$\mu$m & $0.87$ & $0.20$ & $0.22$ & -0.5 to 1.9 & $0.86\pm0.02$ & $0.26\pm0.02$ & $0.20$ & \ref{fig:intscaling} \\
CO~(1-0) vs.\ 22$\mu$m & $0.80$ & $0.01$ & $0.31$ & -0.3 to 2.1 & $0.67\pm0.05$ & $0.23\pm0.05$ & $0.25$ & \ref{fig:intscaling} \\
CO~(1-0)/$12\mu$m vs.\ M$_\star$\tablenotemark{b} & $0.14$ & \nodata & \nodata & 9.0 to 11.25 & $0.13\pm0.05$ & $0.14\pm0.02$ & $0.22$ & \ref{fig:intscaling_mass} \\
CO~(1-0)/$22\mu$m vs.\ M$_\star$\tablenotemark{b} & $0.23$ & \nodata & \nodata & 9.25 to 11.25 & $0.19\pm0.01$ & $-0.06\pm0.01$ & $0.28$ & \ref{fig:intscaling_mass} \\
CO~(1-0)/$12\mu$m vs.\ SFR/M$_\star$\tablenotemark{b,\dag} & $-0.38$ & \nodata & \nodata & -11.5 to -9.25 & $-0.14\pm0.05$ & $0.14\pm0.03$ & $0.20$ & \ref{fig:intscaling_ssfr} \\
CO~(1-0)/$22\mu$m vs.\ SFR/M$_\star$\tablenotemark{b,\dag} & $-0.56$ & \nodata & \nodata & -11.5 to -9.25 & $-0.34\pm0.11$ & $-0.13\pm0.07$ & $0.23$ & \ref{fig:intscaling_ssfr} \\
\hline
CO~(2-1) vs.\ 12$\mu$m & $0.92$ & $0.11$ & $0.23$ & -0.5 to 1.5 & $1.02\pm0.03$ & $0.11\pm0.02$ & $0.21$ & \ref{fig:intscaling} \\
CO~(2-1) vs.\ 22$\mu$m & $0.87$ & $-0.09$ & $0.28$ & -0.3 to 1.9 & $0.85\pm0.05$ & $0.01\pm0.03$ & $0.28$ & \ref{fig:intscaling} \\
CO~(2-1)/$12\mu$m vs.\ M$_\star$\tablenotemark{b} & $0.45$ & \nodata & \nodata & 9.0 to 11.0 & $0.16\pm0.07$ & $0.07\pm0.03$ & $0.17$ & \ref{fig:intscaling_mass} \\
CO~(2-1)/$22\mu$m vs.\ M$_\star$\tablenotemark{b} & $0.47$ & \nodata & \nodata & 9.0 to 11.0 & $0.22\pm0.04$ & $-0.15\pm0.02$ & $0.24$ & \ref{fig:intscaling_mass} \\
CO~(2-1)/$12\mu$m vs.\ SFR/M$_\star$\tablenotemark{b,\dag} & $-0.34$ & \nodata & \nodata & -11.0 to -9.25 & $-0.14\pm 0.12$ & $0.09\pm 0.04$ & $0.19$ & \ref{fig:intscaling_ssfr} \\
CO~(2-1)/$22\mu$m vs.\ SFR/M$_\star$\tablenotemark{b,\dag} & $-0.57$ & \nodata & \nodata & -11.25 to -9.25 & $-0.41\pm 0.09$ & $-0.16\pm0.04$ & $0.23$ & \ref{fig:intscaling_ssfr} \\
\hline
\multicolumn{8}{c}{Rings in radial profiles} \\
\hline
CO~(1-0) vs.\ 12$\mu$m & $0.90$ & $0.38$ & $0.29$ & -1.0 to 1.75 & $0.83\pm0.02$ & $0.40\pm0.02$ & $0.17$ & \ref{fig:resscalingwise} \\
$\ldots$ with sample correction\tablenotemark{c} & \nodata & \nodata & \nodata & \nodata & $0.83$ & $0.21$ & \nodata & \\
CO~(1-0) vs.\ 22$\mu$m & $0.86$ & $0.17$ & $0.39$ & -1.0 to 2.25 & $0.65\pm0.05$ & $0.30\pm0.03$ & $0.21$ & \ref{fig:resscalingwise} \\
$\ldots$ with sample correction\tablenotemark{c} & \nodata & \nodata & \nodata & \nodata & $0.65$ & $0.11$ & \nodata & \\
CO~(1-0) vs.\ 8$\mu$m & $0.91$ & $0.23$ & $0.19$ & -1.0 to 1.75 & $0.88\pm0.05$ & $0.29\pm0.04$ & $0.16$ & \ref{fig:resscalingspitzer} \\
$\ldots$ with sample correction\tablenotemark{c} & \nodata & \nodata & \nodata & \nodata & $0.88$ & $0.10$ & \nodata &  \\
CO~(1-0) vs.\ 24$\mu$m & $0.90$ & $0.28$ & $0.39$ & -1.0 to 2.0 & $0.71\pm0.06$ & $0.33\pm0.03$ & $0.19$ & \ref{fig:resscalingspitzer} \\
$\ldots$ with sample correction\tablenotemark{c} & \nodata & \nodata & \nodata & \nodata & $0.71$ & $0.14$ & \nodata &  \\
\hline
CO~(2-1) vs.\ 12$\mu$m & $0.94$ & $0.13$ & $0.22$ & -1.0 to 2.0 & $1.04\pm0.05$ & $0.09\pm0.04$ & $0.19$ &  \ref{fig:resscalingwise} \\
CO~(2-1) vs.\ 22$\mu$m & $0.91$ & $-0.03$ & $0.29$ & -1.0 to 2.25 & $0.91\pm0.05$ & $-0.09\pm0.04$ & $0.23$ & \ref{fig:resscalingwise} \\
CO~(2-1) vs.\ 8$\mu$m & $0.94$ & $-0.05$ & $0.26$ & -1.0 to 1.75 & $1.14\pm0.03$ & $-0.11\pm0.02$ & $0.21$ & \ref{fig:resscalingspitzer} \\
CO~(2-1) vs.\ 24$\mu$m & $0.91$ & $0.01$ & $0.36$ & -1.0 to 2.0 & $0.95\pm0.07$ & $-0.04\pm0.02$ & $0.23$ &  \ref{fig:resscalingspitzer} \\
\hline
\multicolumn{8}{c}{Individual $17''$ regions} \\
\hline
CO~(1-0) vs.\ 12$\mu$m & $0.71$ & \nodata & \nodata & -1.0 to 2.25 & $0.79\pm0.03$ & $0.41\pm0.03$ & $0.19$ & \ref{fig:pixscalingwise} \\
$\ldots$ with sample correction\tablenotemark{c} & \nodata & \nodata & \nodata & \nodata & $0.79$ & $0.22$ & \nodata & \\
CO~(1-0) vs.\ 22$\mu$m & $0.68$ & \nodata & \nodata & -0.5 to 2.75 & $0.59\pm0.05$ & $0.34\pm0.07$ & $0.20$ & \ref{fig:pixscalingwise} \\
$\ldots$ with sample correction\tablenotemark{c} & \nodata & \nodata & \nodata & \nodata & $0.59$ & $0.15$ & \nodata & \\
CO~(1-0) vs.\ 8$\mu$m & $0.63$ & \nodata & \nodata & -0.75 to 1.5 & $0.89\pm0.06$ & $0.21\pm0.03$ & $0.22$ & \ref{fig:pixscalingspitzer} \\
$\ldots$ with sample correction\tablenotemark{c} & \nodata & \nodata & \nodata & \nodata & $0.89$ & $0.02$ & \nodata & \\
CO~(1-0) vs.\ 24$\mu$m & $0.70$ & \nodata & \nodata & -0.75 to 1.5 & $0.73\pm0.05$ & $0.37\pm0.03$ & $0.21$ & \ref{fig:pixscalingspitzer} \\
$\ldots$ with sample correction\tablenotemark{c} & \nodata & \nodata & \nodata & \nodata & $0.73$ & $0.18$ & \nodata & \\
\hline
CO~(2-1) vs.\ 12$\mu$m & $0.87$ & \nodata & \nodata & -1.0 to 2.25 & $0.96\pm0.05$ & $0.04\pm0.04$ & $0.18$ & \ref{fig:pixscalingwise} \\
CO~(2-1) vs.\ 22$\mu$m & $0.82$ & \nodata & \nodata & -0.5 to 2.25 & $0.80\pm0.04$ & $-0.01\pm0.04$ & $0.20$ & \ref{fig:pixscalingwise} \\
CO~(2-1) vs.\ 8$\mu$m & $0.81$ & \nodata & \nodata & -0.75 to 1.5 & $1.17\pm0.08$ & $-0.19\pm0.05$ & $0.21$ & \ref{fig:pixscalingspitzer} \\
CO~(2-1) vs.\ 24$\mu$m & $0.83$ & \nodata & \nodata & -0.75 to 1.75 & $0.90\pm0.06$ & $0.03\pm0.04$ & $0.19$ & \ref{fig:pixscalingspitzer} \\
\enddata
\tablenotetext{b}{Note that fits vs. M$_\star$ have their intercept at $\log_{10} {\rm M}_\star = 10$ and fits vs. SFR/M$_\star$ have their intercept at $\log_{10}$ SFR/M$_\star = -10$.}
\tablenotetext{\dag}{Correlated axes.}
\tablenotemark{c}{The row reports the fit to the resolved data with an offset applied to the intercept to account for the $+0.19$~dex median offset of galaxies with CO~(1-0) mapping from the integrated relations. The direct fits describe our resolved CO~(1-0) data better, but these corrected fits should better describe the general population of massive, star-forming galaxies captured by the galaxy integrated CO~(1-0) data (Figure \ref{fig:sample}). See \S \ref{sec:intprops} for a more detailed description of the correction.}
\tablecomments{See \S \ref{sec:data} for descriptions of data and methods.}
\end{deluxetable*}

Table \ref{tab:sample1} summarizes our data sets and Table \ref{tab:results} reports the main results of our correlation analysis, fitting, and scatter measurements for each data pair. Figures \ref{fig:intscaling} through \ref{fig:pixscalingspitzer} visualize the results for individual CO-mid infrared data pairs. Then Figures \ref{fig:slope} and \ref{fig:normalization} compare the results for different bands while Figures \ref{fig:intscaling_mass} and \ref{fig:intscaling_ssfr} show how our results relate to integrated galaxy properties.

The sample presented in Table \ref{tab:sample1} and the figures has a few key properties. First, the literature is extensive: there have been $\sim$1,700 CO~(1-0) or CO~(2-1) measurements for integrated galaxies, and we derive matched CO and mid-IR measurements for $\sim$2,600 azimuthal averages (``rings''), corresponding to $\sim 1{,}300$ independent measurements of either CO~(1-0) or CO~(2-1) and mid-IR. Similarly, we compile $\sim$62,000 measurements for individual regions ($\sim$15,000 independent measurements) detected in at least one tracer.

Second, as discussed in \S\ref{sec:methods}, Table \ref{tab:sample1} shows that the mid-IR maps are almost always more sensitive than the the CO maps. CO upper limits paired with mid-IR detections are common for all data pairs, while the reverse is not true. CO is almost never detected without associated detected mid-IR emission. 

Third, the samples and sensitivity vary with line and type of measurements. Integrated galaxy CO~(1-0) data remains significantly more common than integrated CO~(2-1) data, thanks to a number of large integrated galaxy surveys \citep[AMIGA, xCOLD~GASS;][respectively]{AMIGA11,XCOLDGASS17} and several large mapping surveys targeting more distant galaxies that only contribute integrated data to this study \citep[CARMA-EDGE, ALMaQUEST;][]{EDGECALIFA17,ALMAQUESTSURVEY20}. On the other hand, while CO~(1-0) and CO~(2-1) surveys have roughly the same amount of resolved data the CO~(2-1) measurements include far fewer upper limits than the CO~(1-0) data. This reflects the excellent sensitivity of recent mm-wave imaging instruments, especially ALMA, at the $\approx 1.3$~mm wavelength of the CO~(2-1) transition. Despite the presence of significant individual nondetections in the azimuthal average and individual region measurements, we note that our broad, simple masking scheme stacks these low signal-to-noise data into significant binned measurements. We discuss this point more in \S \ref{sec:intprops}.

\subsection{Tight $I_{\rm CO}$ vs.\ $I_{\rm MIR}$ relations} \label{sec:scaling}

\begin{figure*}
\centering
\includegraphics[width=0.8\textwidth]{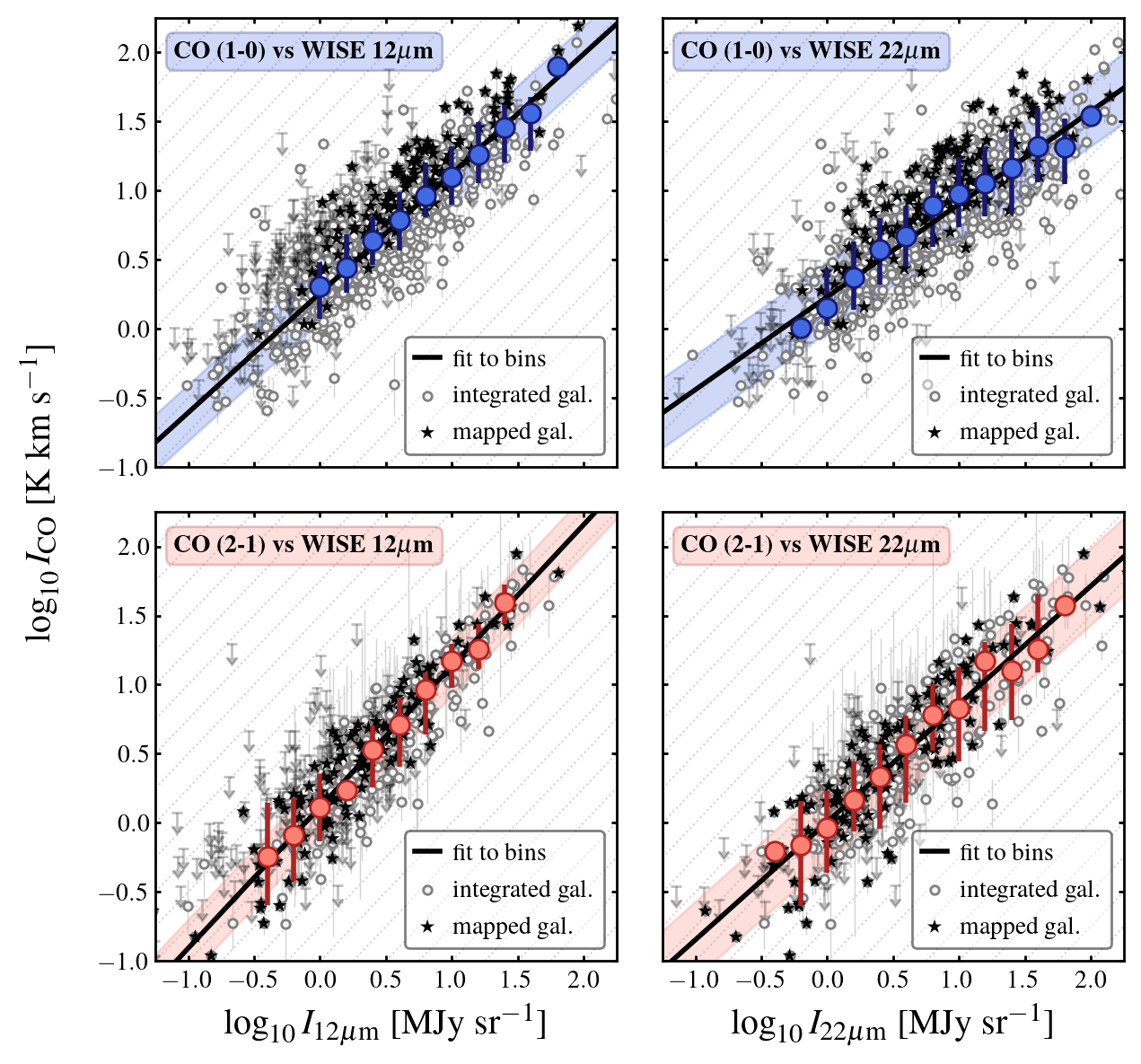}
\caption{\textit{Disk-averaged CO intensity as a function of mid-IR intensity for integrated galaxies.} CO~(1-0) and CO~(2-1) intensity ($y$-axis, top and bottom) as a function of $12\mu$m and $22\mu$m intensity derived from WISE photometry. There are about 800 and 350 galaxies in the CO~(1-0) and CO~(2-1) panels respectively (c.f., Table \ref{tab:sample1}). All measurements are presented as averages within the stellar effective radius. Some of the aperture corrections dominate the size of the error bars, leading to a subset of CO~(2-1) data with larger uncertainties. Black stars show integrated measurements for galaxies that have a resolved CO map. Colored points show the median $I_{\rm CO}$ in bins of mid-IR intensity with error bars indicating the 16{-}84\% range in the data. The light dotted lines in the background all have a slope of $1$, indicating linear relations. The black line and shaded region show a power law fit to the binned data and the rms of the residuals of individual data about that fit. The offset between mapped galaxies and the binned data in CO~(1-0) likely reflects a sample bias (see \S\ref{sec:intprops}).
\label{fig:intscaling}
}
\end{figure*}

Glancing through Table \ref{tab:results} and Figures \ref{fig:intscaling}{-}\ref{fig:pixscalingspitzer} shows our most basic result: for both integrated and resolved galaxies there is a tight relation between $I_{\rm CO}$ and $I_{\rm MIR}$ for all scales and choices of band. For both integrated galaxies and radial profiles, the rank correlation coefficients for the relation of CO to mid-IR are all $\geq 0.80$ and often $>0.9$. The lower correlation coefficients for individual regions simply reflect the lower S/N in those data. The binned relations are well fit by power laws with slopes in the range $\sim 0.7{-}1.2$, i.e., close to linear, with the exact value depending on the mid-IR band and CO line choice. Across our integrated and resolved data sets, the correlations span at least three orders of magnitude in intensity from $I_{\rm MIR} \sim 0.1$~MJy~sr$^{-1}$ to $I_{\rm MIR} \gtrsim 100$~MJy~sr$^{-1}$ and $I_{\rm CO} \sim 0.1$~K~km~s$^{-1}$ to $I_{\rm CO} \gtrsim 100$~K~km~s$^{-1}$. Across this range the scatter in the individual data points about a power law fit is $0.2{-}0.3$~dex (colored regions in Figure~\ref{fig:intscaling}).

This CO-mid-IR relationship is among the tightest observational correlations between distinct spectral bands in extragalactic astronomy. Note that these are relations between ``distance independent'' surface brightnesses and not luminosities.
For luminosities the presence of a correlated distance squared on both axis leads to artificially tighter correlations. Instead we are presenting these relationships in an ``intensive'' way, using $I_{\rm CO}$ and $I_{\rm MIR}$. Had we adopted an extensive quantity like luminosity, their tightness and span would rival the far infrared-HCN relation \citep[e.g.,][]{GAO04SFGAS} or far infrared-radio correlation \citep[e.g.,][]{YUN01RCIR,BELL03SFR}.

The standard interpretation for the CO-mid-IR correlation is that the mid-IR reflects star formation while CO emission captures the fuel for star formation. Though our focus is observational rather than diagnosing the physical emission mechanisms, we note that optically thick CO emission should be brighter in the presence of significant heating and turbulence, while the optical depth of the mid-IR emission will be higher for larger column densities of gas. In that sense, both bands have some element of tracing both star formation and the gas distribution. The tightness and near linearity of the correlation lends circumstantial support to the idea that mid-IR and CO often trace closely related parameters. Over the next few years, the ability of JWST and ALMA to resolve galaxies into discrete regions of recent star formation and abundant molecular gas should yield significant insight on this topic \citep[see, e.g.,][in the PHANGS--JWST first results issue]{KIM22JWST,LEROY22JWSTALMAMUSE,SANDSTROM22JWSTDIFFUSE}.

\begin{figure*}
\centering
\includegraphics[width=0.8\textwidth]{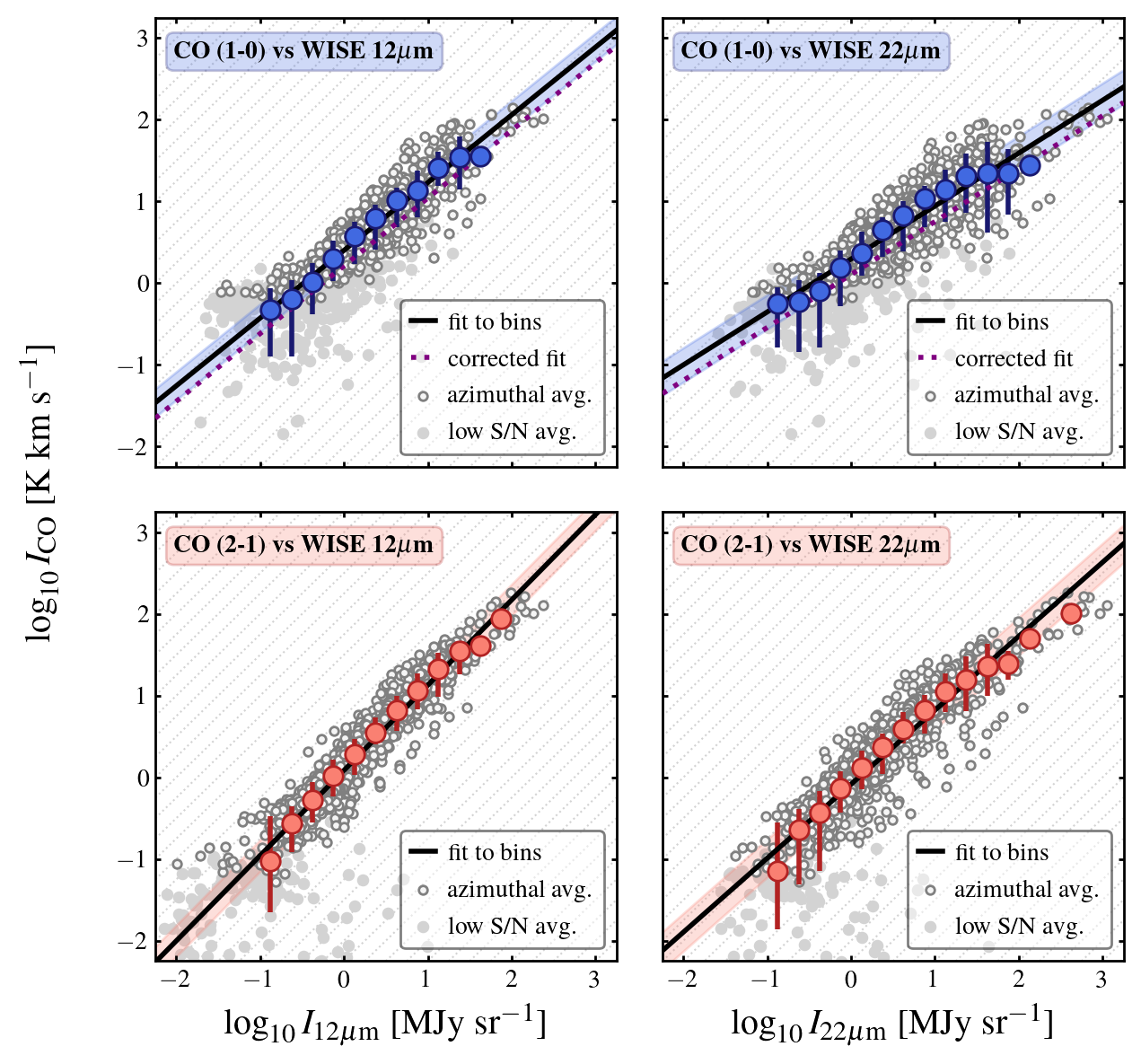}
\caption{\textit{Azimuthally averaged CO intensity as a function of mid-IR intensity at $12\mu$m and $22\mu$m from WISE for moderately resolved galaxies.} As Figure \ref{fig:intscaling} but now showing data for azimuthally averaged rings (radial profiles), within galaxies that have matched-resolution CO and mid-IR mapping (about $1{,}000-1{,}100$ rings per panel detected in box axes depending on the bands, c.f. Table~\ref{tab:sample1}). The light gray points show data with $S/N < 5$, and the binned medians are calculated including these data as well as data with $I_{\rm CO} < 0$, effectively stacking all available maps (about $160-330$ limits per panel depending on the bands). For CO~(1-0) the dotted line shows the expected relation applying a statistical correction (see \S\ref{sec:intprops}) to adjust the resolved mapping sample to better reflect the integrated galaxy sample.
\label{fig:resscalingwise}
}
\end{figure*}

\begin{figure*}
\centering
\includegraphics[width=0.8\textwidth]{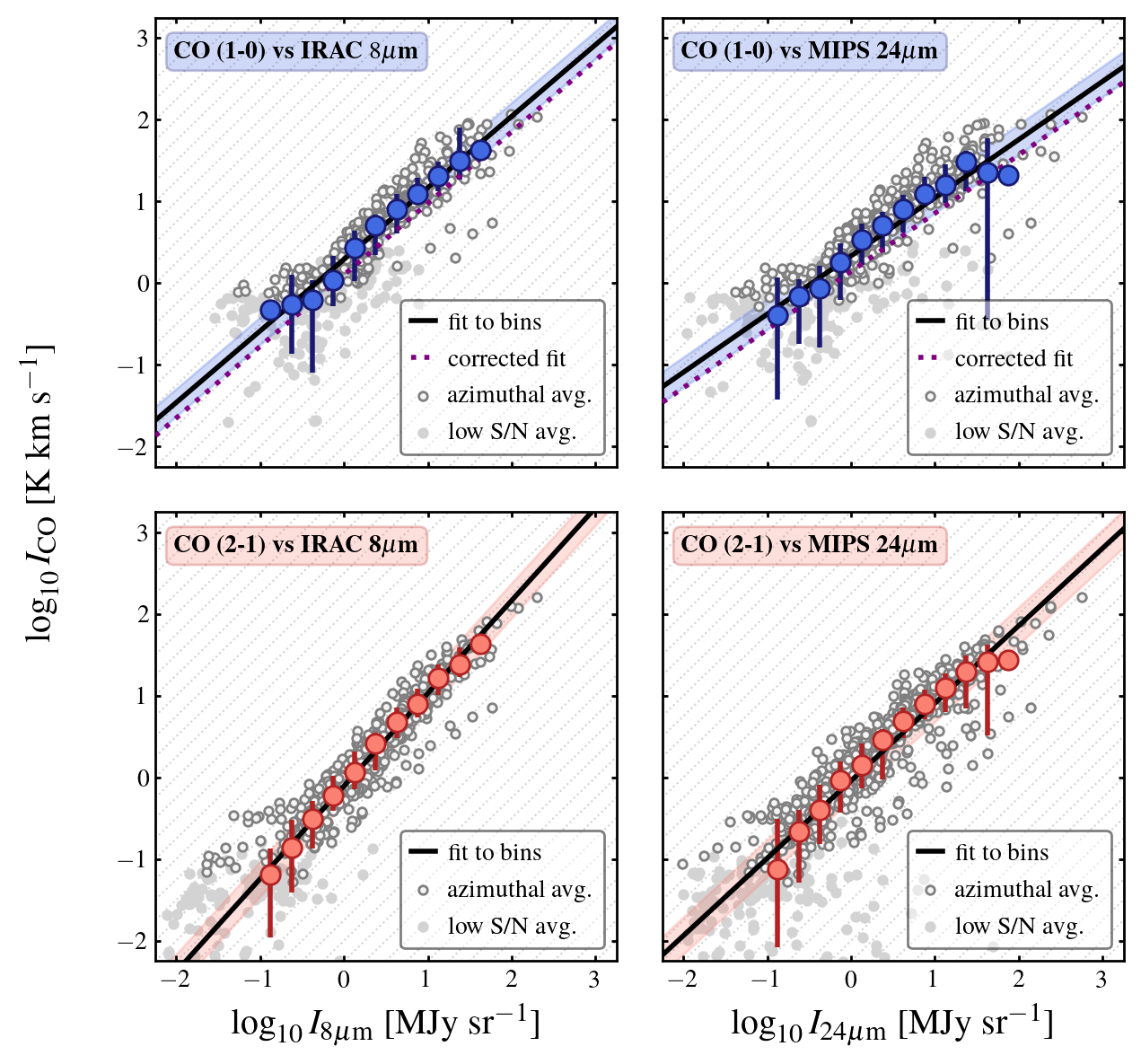}
\caption{\textit{Azimuthally averaged CO intensity as a function of mid-IR intensity at $8\mu$m and $24\mu$m from \textit{Spitzer} for moderately resolved galaxies.} As Figure \ref{fig:resscalingwise} but now showing data for azimuthal averages within galaxies that have mid-IR mapping from \textit{Spitzer}. The overall consistency with the WISE results is good.
\label{fig:resscalingspitzer}
}
\end{figure*}

\begin{figure*}
\centering
\includegraphics[width=0.8\textwidth]{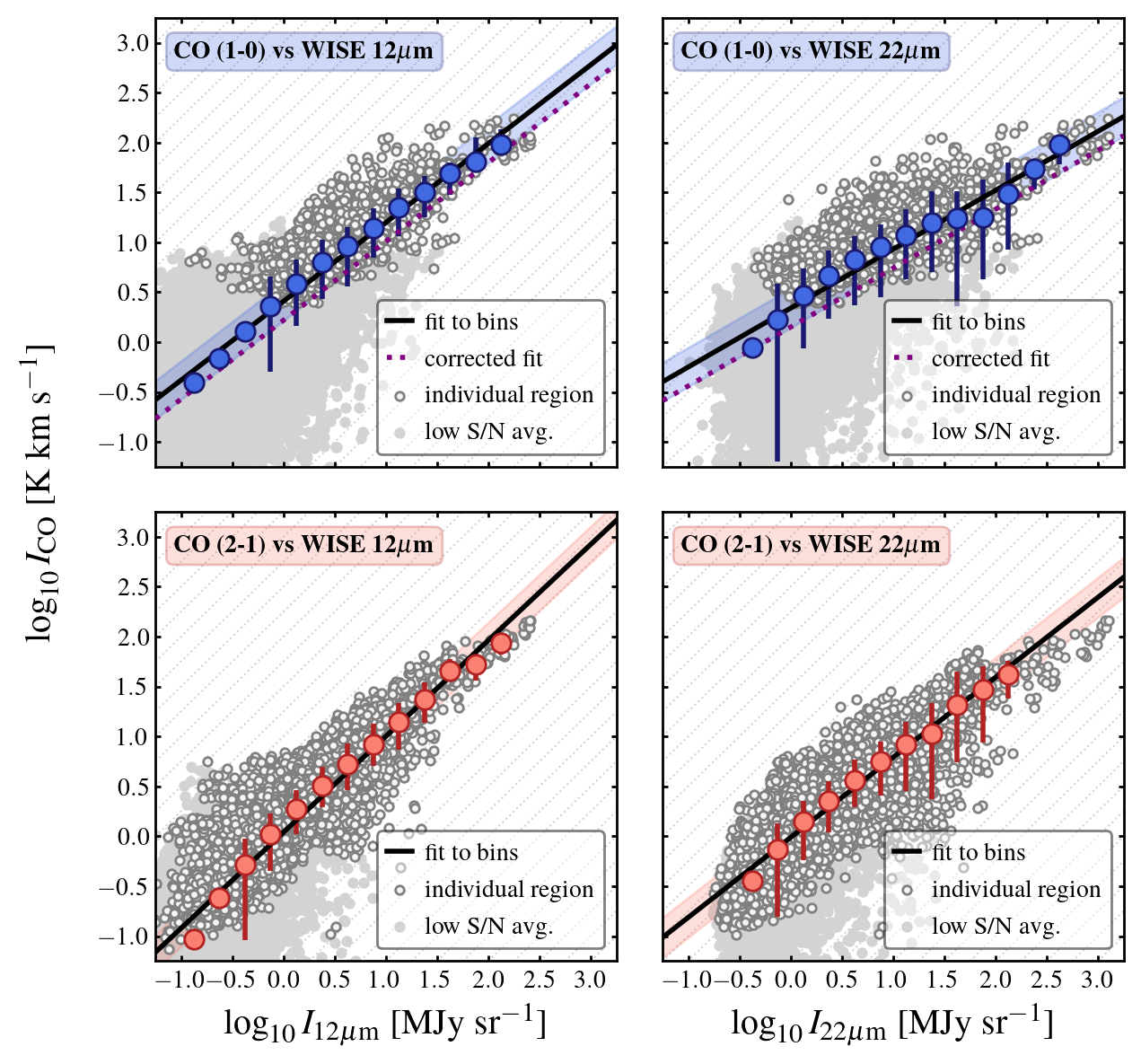}
\caption{\textit{Region-by-region averaged CO intensity as a function of mid-IR intensity at $12\mu$m and $22\mu$m from WISE for moderately resolved galaxies.} As Figure \ref{fig:resscalingwise} but now showing data for individual 17\arcsec\ regions within galaxies that have matched resolution CO and mid-IR mapping. As in Figure \ref{fig:resscalingwise}, the light gray points show data with $S/N < 5$ and the bins include these data as well as data with $I_{\rm CO} < 0$. Given the lower S/N of individual regions, including all data in the binned medians is key to avoid biasing the results. For CO~(1-0) the dotted line shows the expected relation after applying the statistical correction to adjust the resolved mapping sample to better reflect the integrated galaxy sample (\S\ref{sec:intprops}).
\label{fig:pixscalingwise}
}
\end{figure*}

\begin{figure*}
\centering
\includegraphics[width=0.8\textwidth]{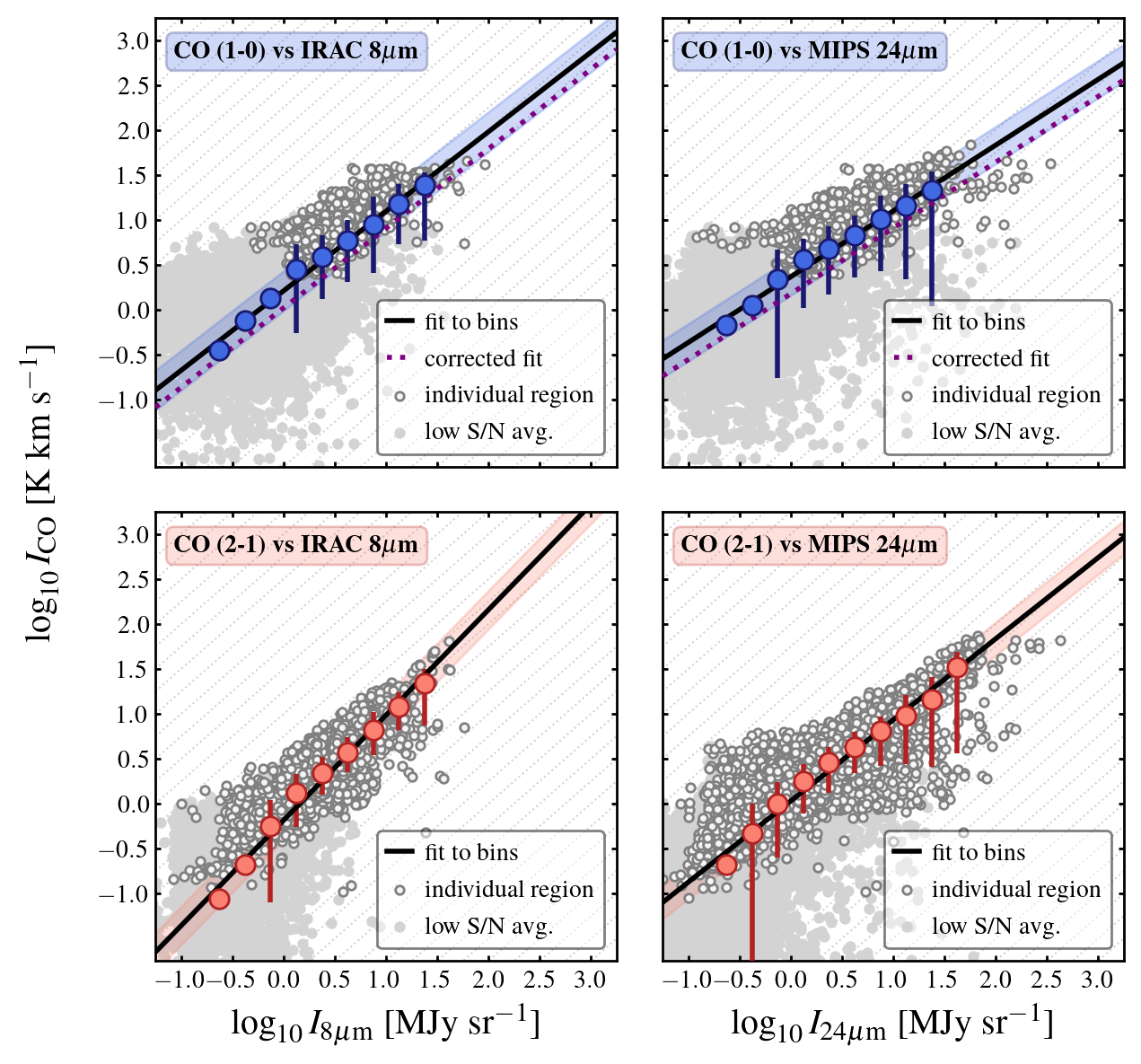}
\caption{\textit{Region-by-region CO intensity as a function of mid-IR intensity at $8\mu$m and $24\mu$m from \textit{Spitzer} for moderately resolved galaxies.} As Figure \ref{fig:pixscalingwise} but now showing data for individual regions within galaxies that have targeted mid-IR mapping from \textit{Spitzer}.
\label{fig:pixscalingspitzer}
}
\end{figure*}

\subsection{Approximately linear slopes for the CO to mid-IR relations}
\label{sec:slope}

\begin{figure}[t!]
\centering
\includegraphics[width=0.45\textwidth]{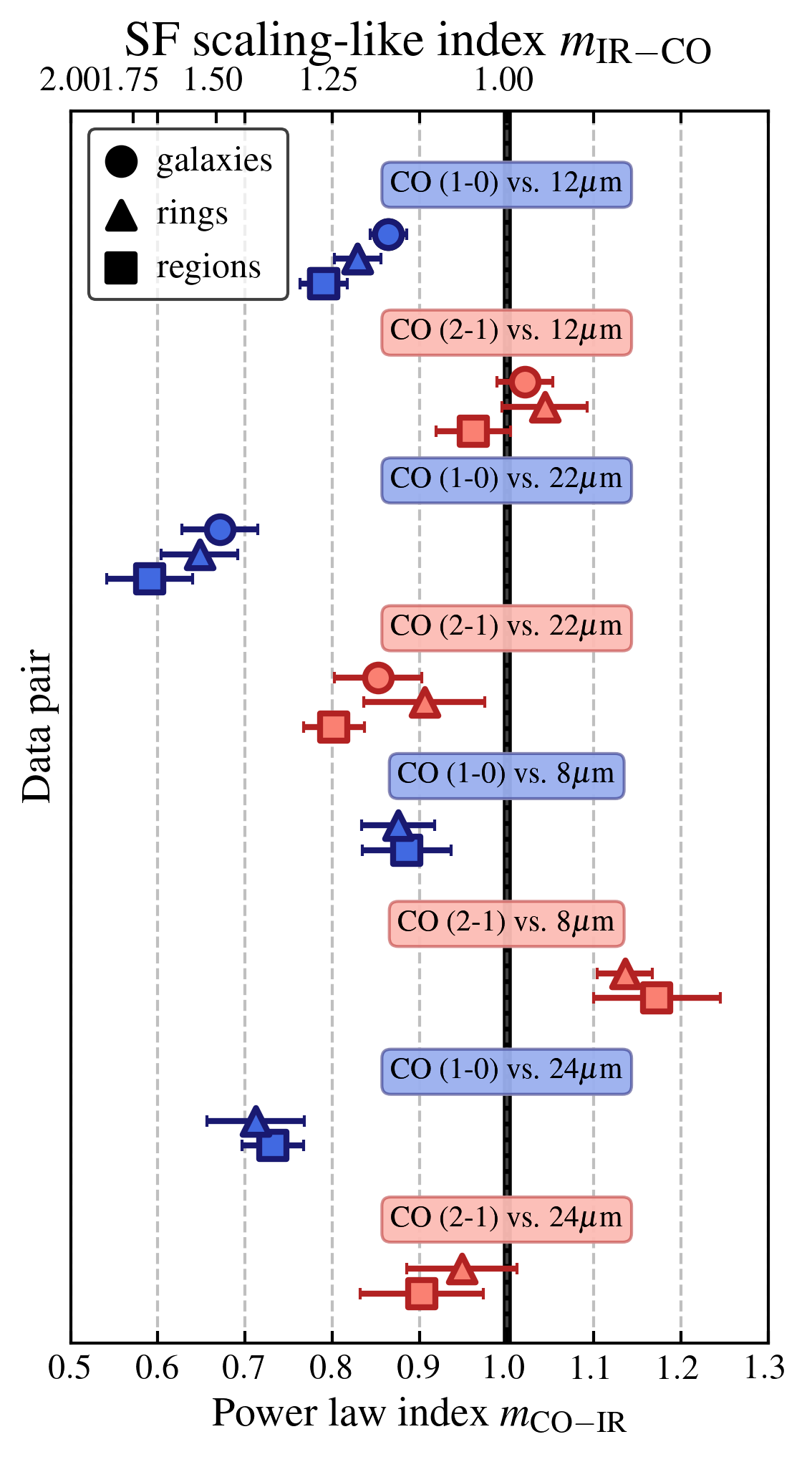}
\caption{\textit{Index of power law fits of $I_{\rm CO}$ vs.\ $I_{\rm MIR}$ for different band and line combinations.} Each point shows the power law index, $m_{\rm CO-IR}$, of a fit relating $I_{\rm CO}$ to $I_{\rm MIR}$ for one line-band combination. We plot results for galaxies, rings, and individual regions separately. The estimated methodological (systematic) error in $m_{\rm CO-IR}$ is $\pm0.1$, the formal errors are very small. The top axis shows the inverse of index, $m_{\rm IR-CO}$, which is the more common sense used to phrase star formation-gas scaling relations. Most results cluster relatively near a slope of $\sim 1$. CO~(2-1) systematically tends to yield a moderately steeper slope than CO~(1-0) for the same mid-IR band. The $8\mu$m and $12\mu$m bands, which contain major PAH features, tend to yield steeper CO to IR slopes than the $22\mu$m and $24\mu$m bands, which capture mostly continuum emission.
\label{fig:slope}}
\end{figure}

Table \ref{tab:results} reports our best fit CO to IR slope for each CO line-IR band pair, and we visualize these together in Figure \ref{fig:slope}. For the most part the results for integrated galaxies, azimuthal averages, and individual regions agree well for a given line-band combination, though there appears to be a mild tendency for the individual region data to show a slightly shallower slope. Given that the integrated galaxy data sets and the resolved data sets are quite distinct, the good overall agreement gives us confidence that we indeed access the slope of an underlying scaling relation that describes local galaxies fairly well.

There are clear patterns within the slopes for different band-line combinations (Figure \ref{fig:slope}). For a given IR band, CO~(2-1) shows a consistently steeper slope than CO~(1-0), which is evident in the offset between red and blue points in Figure \ref{fig:slope}. Averaging over all bands yields a median offset in slope of $\approx +0.20$ from the CO~(1-0) relation to the CO~(2-1) relation with the same mid-IR band. Several recent studies have found that for local star-forming galaxies the CO~(2-1) to CO~(1-0) line ratio varies with the local star formation rate surface density or IR luminosity density as approximately $R_{21} \equiv I_{\rm CO}^{2-1} / I_{\rm CO}^{1-0} \propto \Sigma_{\rm SFR}^{0.15}$ \citep{DENBROK21,YAJIMA21,LEROY22,EGUSA22SFGAS}. Since the mid-IR is often used to trace $\Sigma_{\rm SFR}$ linearly (c.f., Equation \ref{eq:sfrmidir}), our finding agrees well with these results. In essence, our statistical measurement shows that $R_{\rm 21} \propto I_{\rm MIR}^{0.20} \sim \Sigma_{\rm SFR}^{0.20}$ for a very large sample of local galaxies, complementing the more direct studies of small samples or individual galaxies listed above.

We also see offsets among the fitting results for different mid-IR bands. For a fixed choice of line, $12\mu$m shows a steeper $m_{\rm CO-IR}$ slope than $22\mu$m and $8\mu$m shows a steeper $m_{\rm CO-IR}$ slope than $24\mu$m. Across all lines and samples, the median difference between $m_{\rm CO-IR}$ for $12\mu$m and that for $22\mu$m is $+0.18$ and the slope difference between $8\mu$m and $24\mu$m is also $+0.18$. The $12\mu$m and $8\mu$m bands both contain a significant (often dominant) contribution from PAH band features\footnote{The \textit{Spitzer} 8$\mu$m band is almost entirely dominated by the 7.7$\mu$m feature \citep{SMITH07DUST} while \citet{WHITCOMB22MIRCO} find that the wide WISE $12\mu$m band includes up to $\sim 50\%$ contribution from PAHs based on spectroscopy of SINGS \textsc{Hii} regions.}, while the $22\mu$m and $24\mu$m intensities represent mostly continuum emission. This suggests that the PAH features contribute to systematically steeper CO-MIR slopes than the MIR continuum \citep[a finding in very good agreement with spectroscopy-based results from][]{WHITCOMB22MIRCO}.

A possible (albeit non-unique) explanation for this effect would be that the intensity of the PAH emission traces the interstellar medium more directly, including perhaps non-CO emitting gas, while the slightly longer wavelength continuum emission tracks star formation better. Indeed, high spatial resolution imaging in the Milky Way shows that the hot dust 24 $\mu$m emission appears directly related to star formation, while the diffuse PAH emission usually surrounds the star forming region \citep[e.g.,][]{Watson2008}. A similar explanation has been proposed by \citet{CHOWN21SFGAS} and \citet{WHITCOMB22MIRCO} and we might detect statistical evidence for this scenario here. Note however, that differences in how the different mid-IR bands trace the ISM and star formation is likely to be one of degree, with all bands tracing heating by recent star formation or gas to one degree or another. At intermediate scales, PAH-dominated bands still correlate with tracers of star formation and diffuse, stochastically heated dust can still emit mid-IR continuum emission \citep[see][]{LEROY22JWSTALMAMUSE}.

The sense of a steeper CO-MIR slope is that there will be less CO relative to mid-IR emission in the faint, diffuse parts of our data set. This could reflect that lower surface brightness regions within galaxies are expected to hold a larger fraction of their gas in atomic form and perhaps to have a larger fraction of their molecular gas in a CO-dark phase \citep[e.g., see reviews in][]{BOLATTO13REVIEW,SAINTONGEREVIEW22}. These regions are very faint in CO, but will still host PAH emission \citep[e.g.,][]{CHASTENET19DUST,SANDSTROM22JWSTDIFFUSE}. 

%Meanwhile, at least in low metallicity systems, the PAHs abundance appears to be higher in high column density gas \citep[][]{SANDSTROM10}. More, in highly resolved observations bright PAH emission is closely related to the photodissociated surfaces of CO-bright molecular clouds \citep{Wolfire2022}. All of these effects may compound to produce a steeper CO-PAH than CO-continuum relation. 

As discussed in \S \ref{sec:intro} and \ref{sec:methods}, because of the widespread use of the mid-IR in star formation rate estimates, the slopes that we measure relate closely to the literature on star formation scaling relations \citep[e.g.,][among many others all rely heavily on $24\mu$m data]{KENNICUTT07SFR,BIGIEL08SFGAS,KENNICUTT12REVIEW,LEROY13SFGAS}. Our measurements only relate CO and mid-IR emission. They do not include any term accounting for unobscured UV or H$\alpha$ emission or any correction for ``cirrus'' contributions to the mid-IR (i.e., IR emission due to heating from older stellar populations) nor do we consider variations in the CO-to-H$_2$ conversion factor, $\alpha_{\rm CO}$. They are thus not physical estimates of the $\Sigma_{\rm gas}-\Sigma_{\rm SFR}$ relation but direct observables. 

What about using CO as a predictor of mid-IR? Using Equation \ref{eq:irco}, our measured slopes correspond to $m_{\rm IR-CO} \sim 0.9{-}1.4$. For a given IR band, $m_{\rm IR-CO}$ is on average $0.27$ lower for CO~(2-1) compared to CO~(1-0). As discussed above, the PAH-bearing bands will yield shallower MIR vs.\ CO slopes, with $12\mu$m slopes $m_{\rm IR-CO}$ on average $0.25$ lower than those for $22\mu$m.

Finally, we remark that the relationship between $12\mu$m and CO~(2-1) appears nearly linear, with slope $m_{\rm CO-IR}$ almost exactly unity and a rank correlation coefficient $>0.9$, in both our integrated galaxy sample and our radial profile sample. In analyzing PHANGS--ALMA, \citet{PHANGSALMA21} identified the $12\mu$m as the band best correlated with the CO~(2-1) out of several tracers of stellar mass and star formation. The very tight observational relationship appears to hold both broadly and for local star-forming galaxies. Therefore $12\mu$m emission appears to offer the best estimator of the CO~(2-1) brightness of a galaxy or part of a galaxy.

\subsection{Normalizations of the CO to mid-IR relations}
\label{sec:normalization}

\begin{figure}[t!]
\centering
\includegraphics[width=0.45\textwidth]{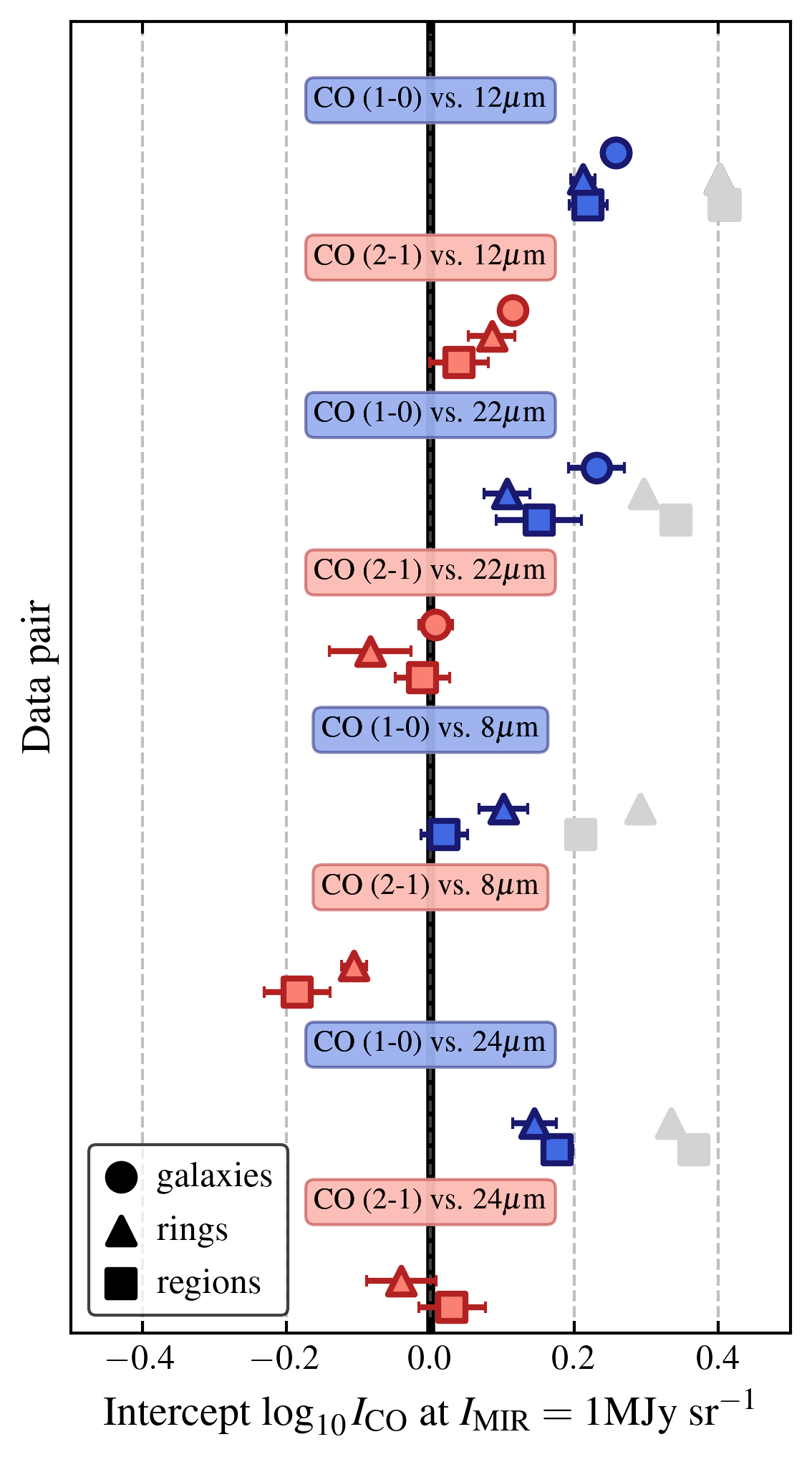}
\caption{\textit{Normalization of power law fits of $I_{\rm CO}$ vs.\ $I_{\rm MIR}$ for different band and line combinations.} As Figure \ref{fig:slope} but now showing the normalization of the best fit power law relations for each line-band combination from Table \ref{tab:results}. The black line marks the simple rule of thumb $I_{\rm CO} \sim 1$~K~km~s$^{-1}$ at $I_{\rm MIR} \sim 1$~MJy~sr$^{-1}$. For CO~(1-0) points we color our best estimates, which include a statistical correction to reflect possible biases in sample selection. However, we also show the original measurements in gray. Overall, we find fainter CO~(2-1) compared to CO~(1-0) at fixed $I_{\rm MIR}$ and the CO-to-mid-IR ratio varies by band with the sense that $I_{\rm CO}/I_{\rm 12\mu m} > I_{\rm CO}/I_{\rm 22\mu m} \sim I_{\rm CO}/I_{\rm 24\mu m} > I_{\rm CO}/I_{\rm 8\mu m}$.
\label{fig:normalization}
}
\end{figure}

We determine the normalization of the $I_{\rm CO}{-}I_{\rm MIR}$ relation, which we report as $b_{\rm CO-IR}$ in Table \ref{tab:results} and compare across bands and transitions in Figure \ref{fig:normalization}. We see again overall similarity, but also real differences as we vary the choice of line and band.

Before examining the differences, we note that Figure \ref{fig:normalization} implies a useful zeroth order approximation that can be used for quick estimates, 

\begin{equation}
\label{eq:ruleofthumb}
\frac{I_{\rm CO}}{I_{\rm MIR}} \sim \frac{\rm 1~K~km~s^{-1}}{\rm 1~MJy~sr^{-1}}~.
\end{equation}

\noindent Given that the slopes we derive are all roughly near unity, this is a useful way to estimate CO from mid-IR intensity or vice versa. This rule of thumb is illustrated as a vertical black line in Figure \ref{fig:normalization}, which shows that it can be expected to work at the factor of $\sim 2$ level. For detailed work one should use the more accurate relations for specific line-band combinations in Table \ref{tab:results}. 

In more detail, for a given IR band, we find the normalization, $b_{\rm CO-IR}$, to be higher for CO~(1-0) than CO~(2-1). This is even more noticeable for the galaxies resolved in CO~(1-0) (e.g., black stars in Figure \ref{fig:intscaling}). As we discuss in \S\ref{sec:intprops}, it seems likely that the root cause of that feature is that the local galaxy CO~(1-0) mapping surveys are biased towards CO-bright galaxies. After accounting for this bias as we describe in \S\ref{sec:intprops}, we find $b_{\rm CO-IR}$ for CO~(1-0) to be a median $0.18$~dex higher than $b_{\rm CO-IR}$ for CO~(2-1) and the same band. The inverse of this factor, $0.65$, closely resembles the typical $R_{21}$ line ratio measured for nearby star-forming galaxies; $R_{21} \approx 0.6{-}0.7$ in recent works, spanning range from $\approx 0.5{-}0.85$ \citep[e.g.,][]{DENBROK21,YAJIMA21,LEROY22}. Thus the offset between lines (either after correcting the resolved CO~(1-0) sample or from only using the integrated galaxies) appears fully consistent with being simply driven by the average CO line ratio (excitation).

The individual IR bands also differ, reflecting their different brightness and consistent with the differences in slopes that we see in \S \ref{sec:slope}. Assuming that these normalizations reflect only the underlying band ratios, this is consistent with $8\mu$m on average being the brightest band studied, showing intensity $\approx 1.3$ times larger than $22\mu$m or $24\mu$m, and $\approx 1.6$ times larger than $12\mu$m.

Just as the slope links to common star formation scaling relations, the CO-to-mid-IR ratio reflected by $b_{\rm CO-IR}$ has a close link to the molecular gas depletion time, $\tau_{\rm dep}^{\rm mol} \equiv M_{\rm mol}/{\rm SFR}$. This quantity captures the normalized rate of star formation per unit molecular gas mass, is of broad interest to understanding star formation in galaxies and is often estimated using combinations of data that include CO line and mid-IR emission \citep[e.g.,][]{LEROY08SFGAS,JANOWIECKI17SFR}. We note that following the approximate translations to mass and star formation rate in Equations \ref{eq:mmolco} and \ref{eq:sfrmidir}, this ratio translates to a molecular gas depletion time of $\tau_{\rm dep} \equiv M_{\rm mol}/{\rm SFR} \sim 0.8{-}1.7$~Gyr depending on the exact choice of mid-IR band and CO line. This agrees very well with a wide range of recent estimates of these quantities \citep[e.g., see][]{LEROY13SFGAS,XCOLDGASS17,SAINTONGEREVIEW22}.

We caution that both these normalizations and our measured slopes will show some scale dependence. A number of studies have shown that at high resolution, tracers of star formation and molecular gas in galaxies resolve into distinct distributions. This leads relationships measured on large scales to show increased scatter on small scales \citep[e.g.,][]{Schruba2010,LEROY13SFGAS,Chevance2020,KIM22SFGAS,PAN22SFGAS}. For the case of the mid-IR this ``breakdown'' in the low resolution scaling relation appears to be somewhat tempered by the fact that mid-IR emission can act as both a star formation and a gas tracer. In first results comparing PHANGS--JWST \citep{PHANGSJWST22} to PHANGS--ALMA and PHANGS--MUSE \citep{PHANGSMUSE22}, \citet{LEROY22JWSTALMAMUSE} compare CO and H$\alpha$ to mid-IR emission on $\sim 70{-}160$~pc scales. Both CO and H$\alpha$ exhibit strong relationships with the mid-IR at these scales, even though the relationship between CO and H$\alpha$ shows clear signs of breaking down. As a result, separate relationships appear to relate CO to mid-IR and H$\alpha$ to mid-IR at high resolution. Analyzing spectroscopic mid-IR data from SINGS, \citet{WHITCOMB22MIRCO} come to a similar conclusion, that the mid-IR acts as both a gas and a SFR tracer to varying degrees. The relationships that we find here still changes at high resolution, but rather than breaking down entirely it appears to resolve into these two separate relationships. CO does still correlate with the mid-IR, simply with a distinct functional relationship and a higher CO-to-mid-IR ratio \citep[see][]{LEROY22JWSTALMAMUSE}.

\subsection{Dependence of the CO to mid-IR ratio on integrated galaxy properties} \label{sec:intprops}

\begin{figure*}
\centering
\includegraphics[width=0.8\textwidth]{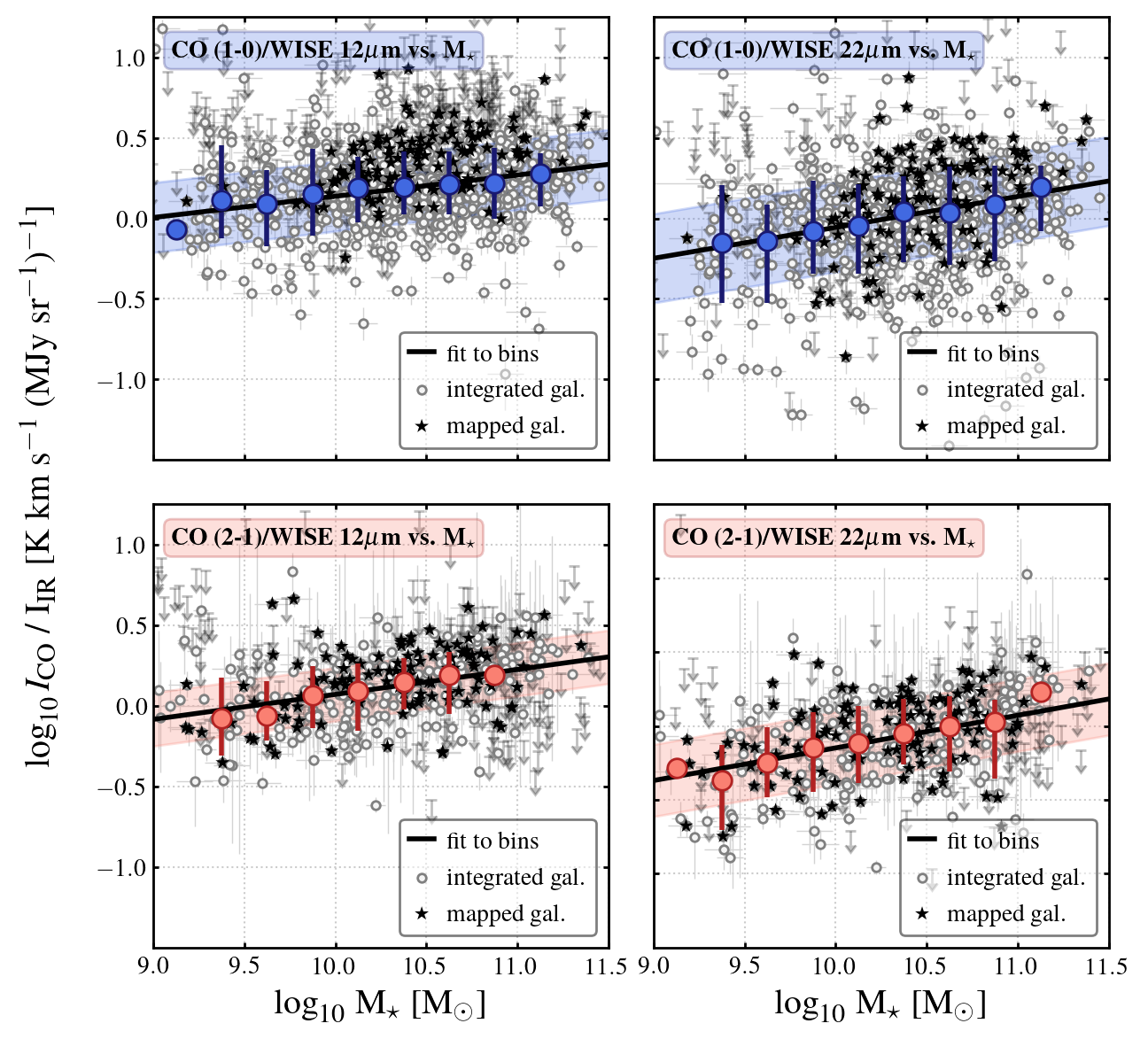}
\caption{\textit{Disk-averaged CO to mid-IR intensity ratio as a function of galaxy stellar mass.} Each point shows a measurement for an integrated galaxy, with galaxies that have a resolved CO map suitable for comparison to the mid-IR (i.e., relatively nearby galaxies with large angular sizes of several arcminutes) marked with black star symbols. The bins show an overall weak correlation of the CO to mid-IR ratio with stellar mass. This could be indicative of a combination of atomic gas and CO-dark molecular gas contributing some mid-IR emission in low mass galaxies, or alternatively comparatively brighter CO emission per molecular mass in high mass galaxies, or perhaps a lower star formation efficiency per unit molecular gas mass at high stellar masses. The set of very nearby galaxies with large CO~(1-0) maps shows significant bias relative to the overall galaxy population sampled by the integrated measurements, very likely reflecting a sample bias since these surveys were optimized to provide high S/N detections.
\label{fig:intscaling_mass}
}
\end{figure*}

\begin{figure*}
\centering
\includegraphics[width=0.8\textwidth]{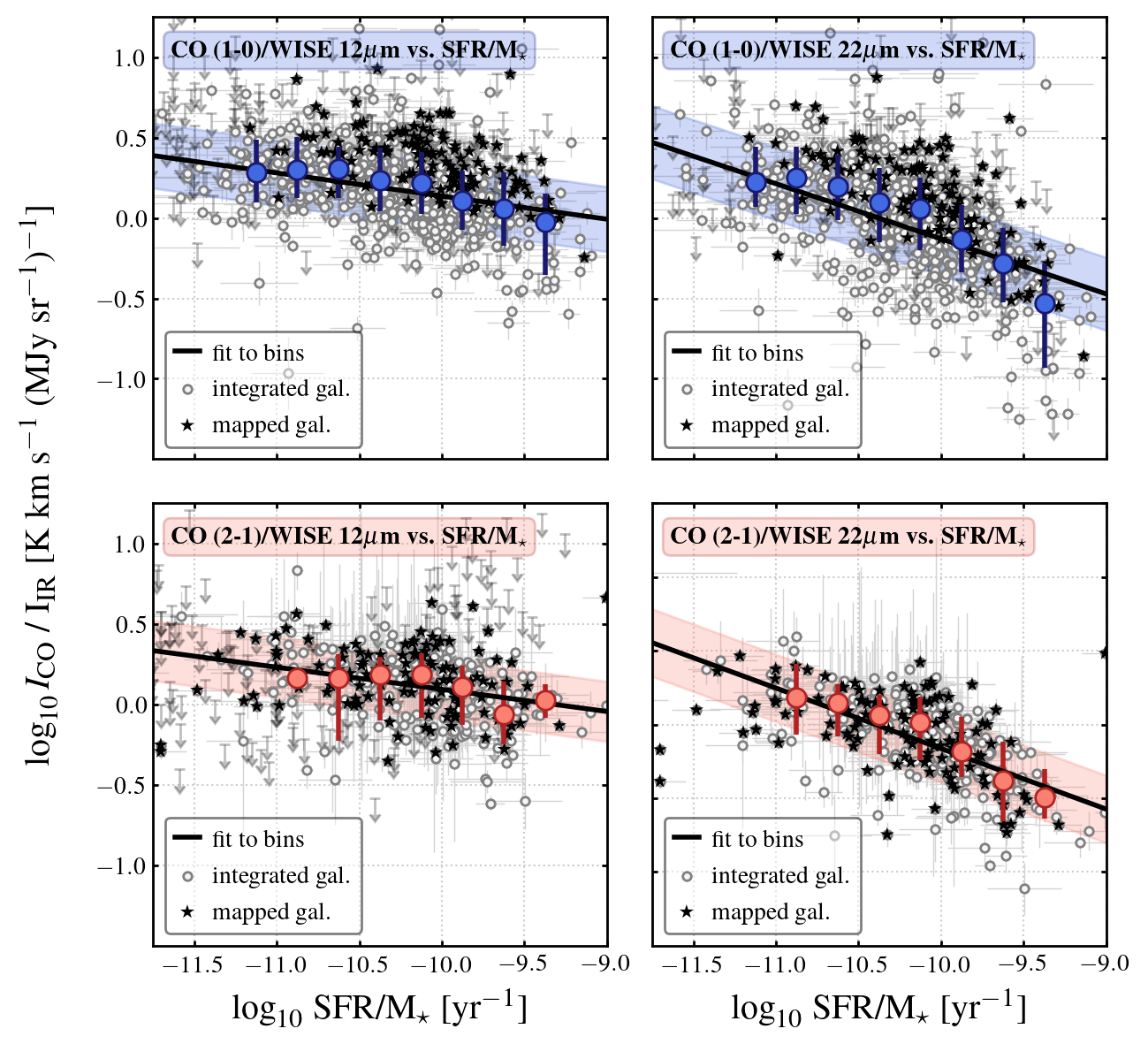}
\caption{\textit{Disk-averaged CO-to-mid IR intensity ratio as a function of galaxy specific star formation rate.} As Figure \ref{fig:intscaling_mass} but now comparing $I_{\rm CO}/I_{\rm MIR}$ to SFR/$M_\star$. We caution that because of the use of $22\mu$m emission as part of the SFR estimate the axes are internally anti-correlated. Nonetheless, the observed anti-correlation between $I_{\rm CO}/I_{\rm MIR}$ to SFR/$M_\star$ matches recent results that link the molecular gas depletion time to offsets from the main sequence \citep{SAINTONGE16STACKING,COLOMBO18,EDGECALIFA20}. The steeper slope for $22\mu$m compared to $12\mu$m emission could reflect a dependence of PAH abundance on SFR/$M_\star$, also broadly consistent with recent observations.
\label{fig:intscaling_ssfr}
}
\end{figure*}

We have so far focused on the intensity-intensity scaling, but the star formation and molecular gas in galaxies also relates to integrated mass, metallicity, and specific star formation rate of the system \citep[e.g., see reviews in][]{TACCONIHIGHZREVIEW20,SAINTONGEREVIEW22}. In Figures \ref{fig:intscaling_mass} and \ref{fig:intscaling_ssfr} we show the CO-to-mid-IR ratio as a function of $M_\star$ and SFR/$M_\star$. Table \ref{tab:results} reports the results of these comparisons along with fits to the intensity scaling relations.

The plots and table show a weak but clear dependence of the CO to mid-IR ratio on stellar mass. The sense of the correlation is that there is more CO emission relative to mid-IR emission in more massive galaxies. Lower mass galaxies also have lower metallicities and fainter CO emission \citep[see][]{SANCHEZ20REVIEW,BOLATTO13REVIEW, SAINTONGEREVIEW22}. In principle, both the dust producing mid-IR emission and the molecular gas producing CO become less abundant at low stellar mass \citep[see][]{BOLATTO13REVIEW,GALLIANO18REVIEW}. The results here suggest that CO is more quickly depressed in low mass galaxies than the mid-IR emission, with a best fit slope against stellar mass of $m \sim 0.2$. 

There are a number of plausible explanations for this trend. The star formation per unit gas mass may be higher in low mass galaxies (and SFR/$M_\star$ certainly anti-correlates with $M_\star$). More massive galaxies may also be more likely to have a significant fraction of their CO emission from bar-fed nuclear starburst regions. The gas in these regions tends to be more emissive due to increased temperature, dynamically driven line widths, and low opacity, which may cause an increase in CO luminosity for a given molecular mass \citep[e.g.,][]{BOLATTO13REVIEW,TENG2021} and lead to a lower CO-to-mid-IR ratio. Alternatively, a larger fraction of mid-IR emission may emerge from dust associated with CO-dark gas and atomic gas in low mass galaxies compared to high mass galaxies. There is no \textit{a priori} physical reason for mid-IR emission to only trace CO-bright molecular gas \citep[e.g.,][]{WALTERBOS87}, so we consider this explanation a likely one. \citet{SANDSTROM22JWSTDIFFUSE} address this issue in the first results from PHANGS--JWST and more future highly resolved analysis using JWST and ALMA should help shed light on this topic.

Figure \ref{fig:intscaling_ssfr} and Table \ref{tab:results} also report the scaling of $I_{\rm CO}/I_{\rm MIR}$ with specific star formation rate, SFR$/M_\star$. We do caution that because our SFR estimates involve $22\mu$m emission the axes in this plot are anti-correlated (this is the only such case in the paper). The plot suggests an overall anti-correlation, especially at $22\mu$m, with the sense that CO drops relative to mid-IR emission as SFR/$M_\star$ increases. Taken at face value, this agrees well with a series of recent results that show that the depletion time of molecular gas anticorrelates with SFR/$M_\star$ and distance to the star forming main sequence \citep[][]{SAINTONGE16STACKING,COLOMBO18,EDGECALIFA20}. In simple terms, molecular gas as traced by CO appears less efficient at forming stars in galaxies with low specific star formation rate and more efficient in starburst galaxies. The anti-correlation of CO-to-mid-IR with SFR/$M_\star$ in Table \ref{tab:results} and Figure \ref{fig:intscaling_ssfr} is consistent with this finding, though a more detailed statistical analysis controlling for the correlated axes would help validate our simple correlation measurement.

Figure \ref{fig:intscaling_ssfr} shows steeper slopes for the CO-to-22$\mu$m ratio correlation with SFR/$M_\star$ than for the CO-to-12$\mu$m ratio. Assuming that this is not merely a chance byproduct of internal correlations, it may reflect that the $12\mu$m to $22\mu$m ratio is varying in response to destruction of PAHs in regions of high specific SFR and intense star formation. This phenomenon is observed within local galaxies \citep[e.g.,][]{CHASTENET19DUST}, and the $12\mu$m-to-22$\mu$m does drop with increasing SFR/$M_\star$ \citep[e.g., see Figure 21 in][]{LEROY19Z0MGS}.

Finally, Figures \ref{fig:intscaling_mass} and \ref{fig:intscaling_ssfr}, along with Figure \ref{fig:intscaling}, allow us to examine the selection function for the local mapping surveys used for the resolved analysis. In these three figures we mark the local CO mapping targets as black stars. The surveys that we use for the integrated CO sample reflect the massive, star-forming galaxy population well by design \citep[][and see \S \ref{sec:data} and Figure \ref{fig:sample}]{XCOLDGASS17,EDGECALIFA17,ALMAQUESTSURVEY20,EDGECALIFA20}, and WISE $12\mu$m and $22\mu$m imaging covers the entire sky, so we expect the correlations arising from those data to suffer the least from sample bias. The figures show that overall, the CO~(2-1) mapping surveys mostly sample the same parameter space as the integrated galaxy measurements. Indeed, the CO~(2-1) sample is heavily dominated by PHANGS--ALMA, which attempted a volume-limited selection of relatively massive galaxies on or near the star forming main sequence and in that sense represents well the population of star-forming galaxies with $\log_{10} M_\star \approx 9.5{-}11$\,M$_\odot$. By contrast, the CO~(1-0) mapping surveys appear systematically offset towards high CO brightness in all three plots. The local CO~(1-0) mapping sample is dominated by COMING, and \citet{COMING19} do explicitly state that CO brightness is a key criterion of the selection, so this bias might reasonably be expected. 

We attempt a first-order correction for this bias by solving for the median offset between the mapped points (i.e., the black stars) and the best-fit line from the full integrated data set in each fit relation in Figures \ref{fig:intscaling}, \ref{fig:intscaling_mass}, and \ref{fig:intscaling_ssfr}. On average, the median offset required to bring the CO~(1-0) mapping sample in line with the full integrated galaxy sample is $\Delta \log_{10} I_{\rm CO} = -0.19$~dex (a factor of 1.55). We have applied this shift as a ``statistical correction'' in Table \ref{tab:results} and Figures \ref{fig:resscalingwise} through \ref{fig:normalization}. If we solve for the analogous quantity for CO~(2-1) we find a median offset of only $0.04$~dex (a factor of 1.10), which we do not apply.

\subsection{Differences between CO~(2-1) and CO~(1-0) and a useful empirical predictor of the line ratio} \label{sec:r21}

\begin{figure}[t!]
\centering
\includegraphics[width=0.45\textwidth]{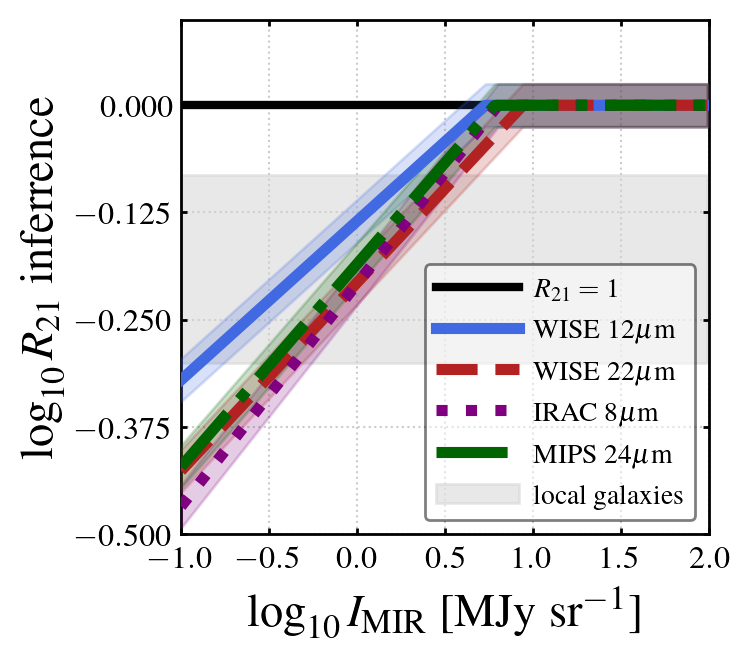}
\caption{\textit{Statistical inference of the CO~(2-1)/CO~(1-0) ratio as a function of mid-IR intensity.} Inferred dependence of the CO~(2-1) to CO~(1-0) line ratio, $R_{21}$, on mid-IR intensity at different bands using our full data set and assuming all surveys sample the same underlying galaxy population (Equation \ref{eq:r21}). The noise and shading are added to allow one to more easily distinguish the lines by eye and we manually cap $R_{21}$ at $1$, which we also recommend when implementing Equation \ref{eq:r21}. The statistical corrections to the local CO~(1-0) mapping surveys have been applied, so that these represent our best estimates of a general relationship applicable to local star-forming galaxies. The gray band shows the observed $16{-}84\%$ range of $R_{21}$ for local galaxies from \citet{LEROY22} and the slope of the inferred relations $R_{21} \propto I_{\rm MIR}^{0.2}$ agrees well with recent results estimating the dependence of $R_{21}$ on $\Sigma_{\rm SFR}$.
\label{fig:r21}
}
\end{figure}

In \S\ref{sec:slope} and \S\ref{sec:normalization} we discussed that the differences in slope and normalization observed when using CO~(1-0) and CO~(2-1) appear to agree well with recent work on the CO~(2-1) to CO~(1-0) line ratio, $R_{21}$. In fact, if we assume that the relations measured for each line accurately reflect the same underlying galaxy population, we can make a much more powerful inference. By dividing the scaling relation for $I_{\rm CO 2-1}$ by that for $I_{\rm CO 1-0}$, we can predict the line ratio as a function of mid-IR intensity. As mentioned above, several recent direct studies of $R_{21}$ have found a clear dependence of the line ratio on tracers of the star formation surface density, resulting in $R_{21} \sim \Sigma_{\rm SFR}^{0.15}$ \citep{DENBROK21,YAJIMA21,LEROY22}. In contrast to these studies, the results here are statistical in nature. On the other hand, given our large sample these relations should  describe the general population of massive, star-forming galaxies near the star-forming main sequence, rather than the smaller samples where this ratio has been studied directly. When relevant, we apply the statistical correction described in \S \ref{sec:intprops} to the CO~(1-0) mapping results.

Figure \ref{fig:r21} visualizes the resulting predictions, which are:

\begin{equation}
\label{eq:r21}
R_{21}\approx\begin{cases} 
0.73~I_{\rm 12\mu m}^{0.19} \\
0.62~I_{\rm 22\mu m}^{0.22} \\
0.62~I_{\rm 8\mu m}^{0.26} \\
0.65~I_{\rm 24\mu m}^{0.24}
\end{cases}
\end{equation}

\noindent with all intensities in MJy~sr$^{-1}$ and the last two relationships being less accurate because they rely only on the much smaller set of targeted \textit{Spitzer} mapping data. As illustrated in Figure \ref{fig:r21}, we recommend capping $R_{21}$ at a value of $R_{21}=1$ when implementing these relationships, which is the expectation for optically thick emission from gas at uniform temperature. As discussed in \citet{LEROY22} there is also a likely plausible lower bound, but the exact value of this is uncertain\footnote{\citet{LEROY22} present an extensive discussion of literature measurements and physical expectations, including model predictions, for ratios among the low-$J$ CO rotational lines and we refer to that paper for background.}.

Although this is a statistical measurement, Equation \ref{eq:r21} should be as robust as any currently available general prescription for $R_{21}$ in the literature.

\subsection{Location of PHANGS-JWST in the moderately resolved scaling relations}

\begin{figure*}[t!]
\centering
\includegraphics[width=0.8\textwidth]{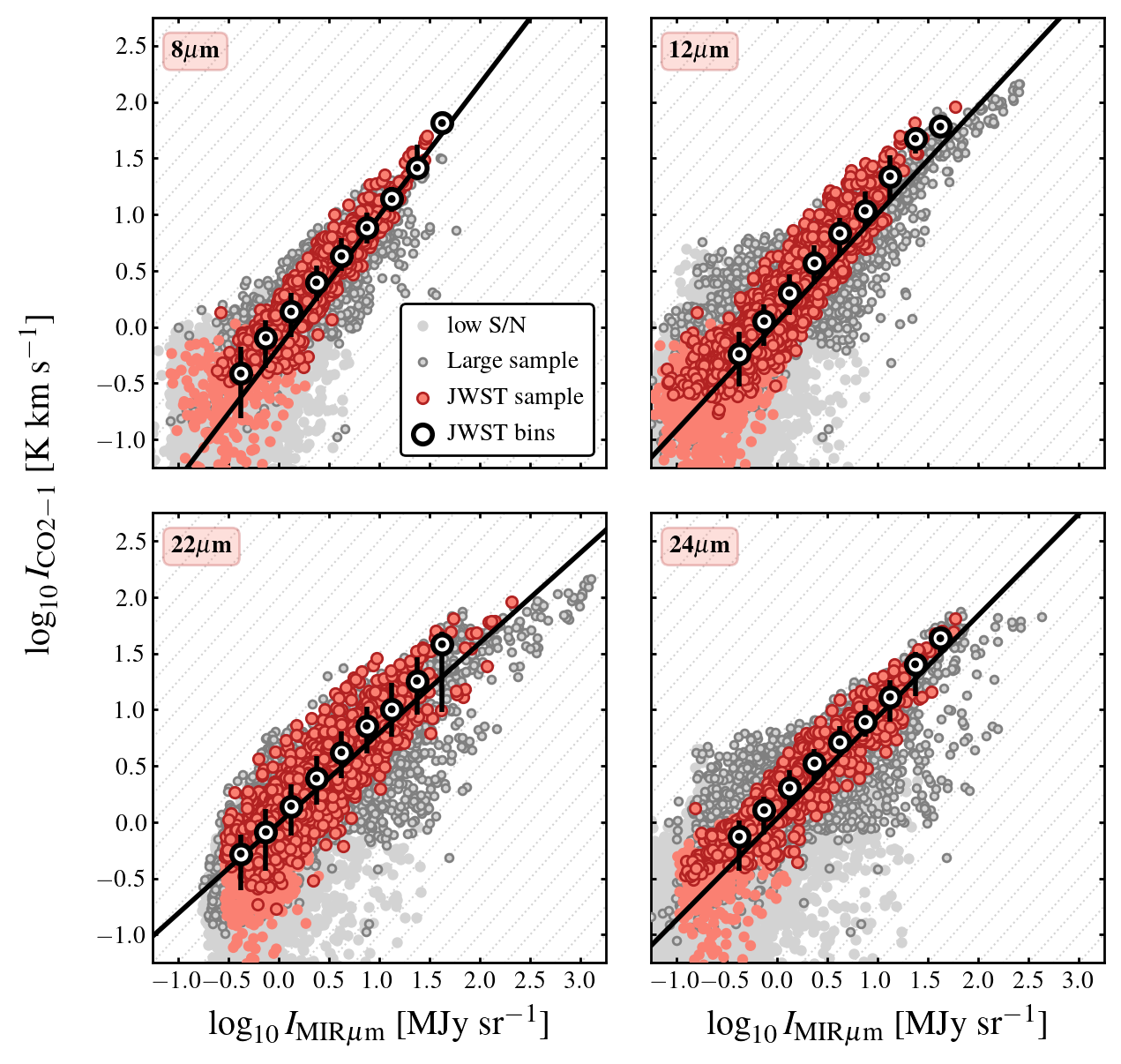}
\caption{\textit{The 19 targets of the full PHANGS-JWST survey relative to our larger sample in $I_{\rm CO}{-}I_{\rm MIR}$ space.} As Figure \ref{fig:pixscalingwise}, but each panel shows our compiled region-by-region data in gray with results, showing only CO~(2-1). Red points show the subset of data for the 19 PHANGS--JWST targets. Though very slightly biased towards regions rich in CO emission, the full PHANGS--JWST survey reflects the larger galaxy population well and shows the same low resolution CO~(2-1) vs.\ mid-IR scalings that we observe in the rest of the local galaxy population.
\label{fig:phangsjwst_sample}
}
\end{figure*}

\begin{figure*}[t!]
\centering
\includegraphics[width=0.8\textwidth]{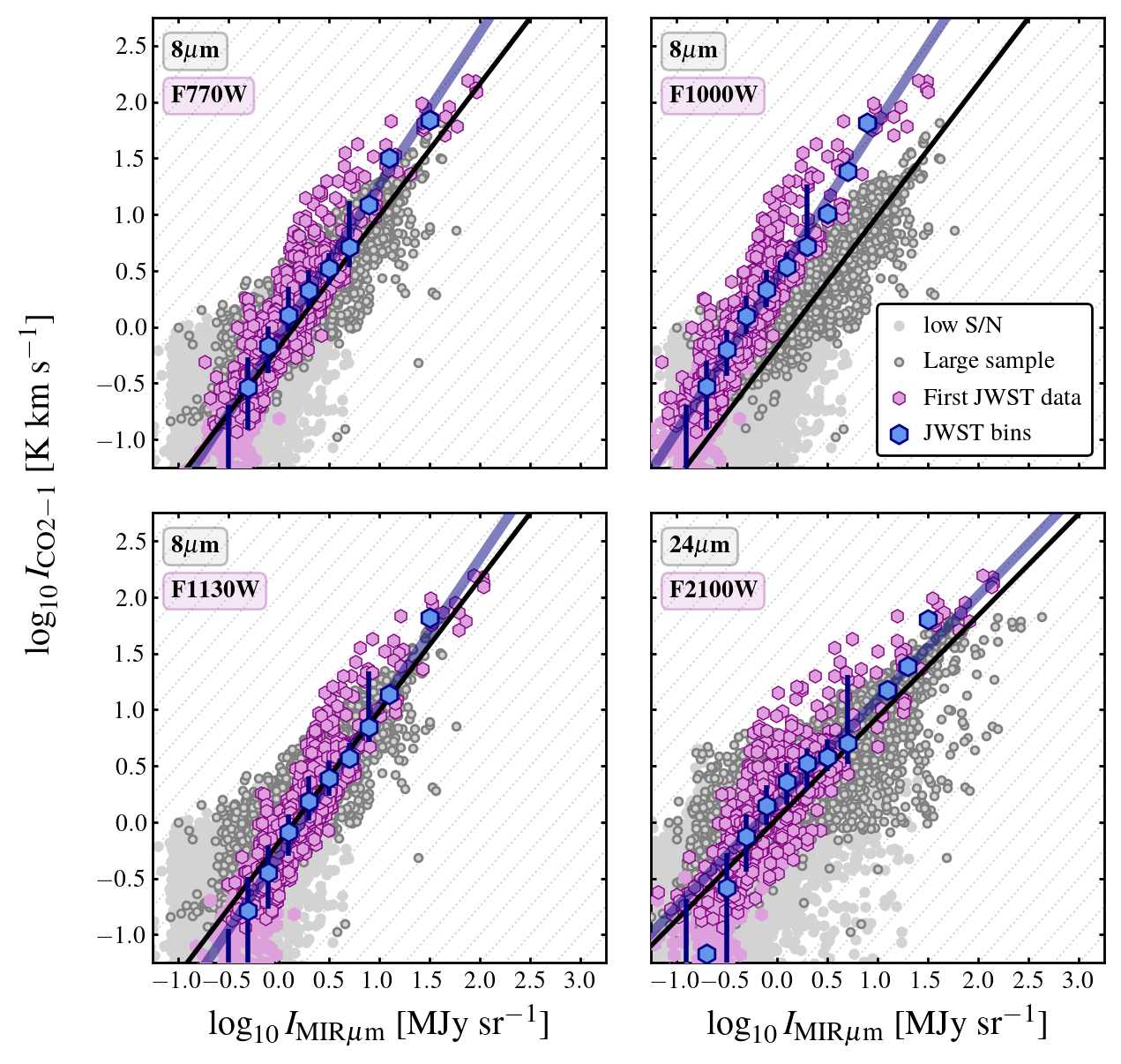}
\caption{\textit{First four PHANGS--JWST target galaxies compared to our larger sample in $I_{\rm CO}{-}I_{\rm MIR}$ space.} As Figure \ref{fig:phangsjwst_sample}, but now each panel shows results from the first four PHANGS--JWST targets in blue and purple. Each panel shows results for a different filter, the blue points show binned versions of the PHANGS--JWST data, and the purple lines shows fits to these binned data. For comparison, we also plot individual region results for our larger dataset in the background of each panel, and show the fit to those data in black. We compare F770W, F1000W, and F1130W to the $8\mu$m band and F2100W to the $24\mu$m band. Given the small sample of four current targets, the agreement in between the PHANGS--JWST fits and those for our larger sample (Table \ref{tab:jwst}) appears excellent. The F770W and F1130W results closely match those for $8\mu$m, while the F1000W fit shows a similar slope offset to indicate the fainter F1000W. F2100W and $24\mu$m also show good agreement.
\label{fig:phangsjwst_data}
}
\end{figure*}

\begin{deluxetable}{l|c|c|c}[ht!]
\tabletypesize{\small}
\tablecaption{Low resolution CO vs. mid-IR relations for PHANGS--JWST \label{tab:jwst}}
\tablewidth{0pt}
\tablehead{
\colhead{CO~(2-1) vs.} & 
\colhead{Bin range} &
\colhead{$m$} &
\colhead{$b$} \\
\colhead{} & 
\colhead{($\log_{10}$ MJy~sr$^{-1}$)} & 
\colhead{} &
\colhead{}
}
\startdata
F770W & -0.4 to 1.2 & $1.36\pm 0.08$ & $-0.10\pm 0.03$ \\
F1000W & -0.8 to 1.0 & $1.37 \pm 0.05$ & $0.44 \pm 0.02$ \\
F1130W & -0.4 to 1.2 & $1.32 \pm 0.09$ & $-0.30 \pm 0.04$ \\
\hline
$8\mu$m\tablenotemark{a} & -0.75 to 1.75 & $1.17 \pm 0.08$ & $-0.19 \pm 0.05$ \\
\hline
F2100W & -0.5 to 1.375 & $0.92 \pm 0.09$ & $0.16 \pm 0.03$ \\
\hline
$24\mu$m\tablenotemark{a} & -0.75 to 1.75 & $0.90 \pm 0.06$ & $0.03 \pm 0.04$ \\
\enddata
\tablenotetext{a}{Reproduced from ``individual $17''$ resolution regions'' results in Table \ref{tab:results} for comparison. Includes all galaxies.}
\tablecomments{Fitting results from binning and fitting individual region measurements in the first four PHANGS--JWST galaxies at $15''$ resolution.}
\end{deluxetable}

During Cycle 1, the PHANGS-JWST Treasury program \citep{PHANGSJWST22} will observe mid-IR emission at $7.7\mu$m (F770W), $10\mu$m (F1000W), $11.3\mu$m (F1130W), and $21\mu$m (F2100W) from $19$ local star-forming galaxies. All of these targets have corresponding high resolution ALMA data \citep{PHANGSALMA21}. In Figure \ref{fig:phangsjwst_sample} we compare the $19$ targets of the PHANGS--JWST survey to our full low resolution data set. The PHANGS--JWST targets all have ALMA CO~(2-1) and WISE $12\mu$m and $22\mu$m data, though only about half have IRAC $8\mu$m and MIPS $24\mu$m data. In the Figure, the red points showing data for PHANGS-JWST targets do a reasonable job of spanning the full range of CO and mid-IR intensities seen in the literature sample and also appear broadly consistent with the CO vs.\ mid-IR relations seen for the broader sample.

At the time of submission of this manuscript, four of the PHANGS--JWST targets have already been observed by the MIRI instrument on JWST (NGC\,628, NGC\,1365, NGC\,7496, and IC\,5332). \citet{PHANGSJWST22} in the PHANGS--JWST First Results issue report details of the survey design, data reduction, and quality assurance of these MIRI data. Briefly, each pointing was observed for $89$ (F770W), $122$ (F1000W), $311$ (F1130W), and $322$ (F2100W) seconds. Reduction mostly followed the standard JWST procedures, with the sky and instrumental background removed by subtracting a nearby ``off'' field. After processing by the JWST pipeline, the individual pointings were mosaicked together and the different MIRI bands were compared to set a common background level. Finally, the background level of all bands was anchored to previous wide-field observations by either \textit{Spitzer} or WISE following the procedure described in \citet{LEROY22JWSTALMAMUSE}. Overall, the flux calibration of the JWST data appear consistent with these previous observations and the background level is uncertain at the $\pm 0.1$~MJy~sr$^{-1}$ level.

In Figure \ref{fig:phangsjwst_data}, we show results from JWST for the first four PHANGS--JWST targets. The purple points show PHANGS--ALMA CO~(2--1) as a function of mid-IR intensity in four JWST filters\footnote{The JWST and ALMA data in purple in Figure \ref{fig:phangsjwst_data} are all matched at $15''$ as described in Appendix A in \citet{LEROY22JWSTALMAMUSE}, almost identical to our common $17\arcsec$ working resolution.}. The blue points show binned versions of the PHANGS--JWST data and the purple line shows a fit to these bins. For comparison, we also plot the data for individual regions from our larger data set. For the F770W, F1000W, and F1130W bands, we show CO~(2-1) as a function of $8\mu$m emission measured by \textit{Spitzer}.

We compare F770W and F1130W, which are expected to be PAH-dominated, to results for the PAH-dominated $8\mu$m band. Empirically F1000W tracks these other two MIRI bands closely, though whether F1000W is PAH dominated is less clear, so we also compare it to $8\mu$m. For F2100W we compare to $24\mu$m emission, because both bands are expected to be continuum dominated at similar wavelengths. The black solid lines show the fits to this larger sample. Table \ref{tab:jwst} reports the results of our fits to the JWST data. It also reproduces the $8\mu$m and $24\mu$m individual region fitting results from Table \ref{tab:results} for ease of comparison.

Table \ref{tab:jwst} and Figure \ref{fig:phangsjwst_data} show good overall agreement between the slopes in the JWST data and the comparison data. The PAH-tracing bands all show slopes $\sim 1.3{-}1.4$, similar to the slope of $\sim 1.2$ measured for the $8\mu$m. Meanwhile the CO as a function of $21\mu$m shows a slope of $\sim 0.9$, similar to that relating CO~(2-1) to $24\mu$m. At the most basic level, we confirm that the scalings captured at $\sim 15'' \sim 1$~kpc by PHANGS--JWST quantitatively match those observed in much larger samples at similar bands using \textit{Spitzer}. At a practical level, this provides a quantitative link between the more focused, high resolution JWST observations and the broader low resolution mid-IR literature. We do caution that so far, we have only analyzed four galaxies, and because of this the data from JWST span a limited intensity range. Moreover, essentially all of the bright CO and mid-IR emission within this small sample come from the inner region of a single galaxy, NGC~1365 \citep[e.g., see][]{SCHINNERER22JWST}. With these caveats in mind, the good agreement between such a small sample of JWST results and extensive previous mid-IR mapping in Figure \ref{fig:phangsjwst_data} may be even more impressive.

Physically, the slope of $\sim 1.2{-}1.4$ for the PAH tracing F770W, F1000W, and F1130W bands indicates that the mid-IR emission appears brighter relative to CO~(2-1) emission at low mid-IR intensity. As discussed in \S \ref{sec:slope}, one plausible interpretation of these slopes would be that mid-IR emission emerges from dust mixed with atomic gas or CO-dark molecular gas or low excitation gas. This suggests, the mid-IR may even act as an effective tracer of gas poorly traced by CO. 

The resolution and sensitivity of JWST make this a potentially exciting prospect because we have relatively few ways to trace non CO-emitting neutral gas at high angular resolution. More validation and calibration are needed to unlock the potential of the mid-IR as a quantitative ISM tracer at high resolution \citep[at low resolution see][]{GAO19MIRCO,CHOWN21SFGAS,GAO22MIRCO}, but this work is proceeding quickly. Here, at low resolution we have verified the quantitative link between the PHANGS--JWST results and a larger sample. In this issue, \citet{SANDSTROM22JWSTDIFFUSE} carefully examine low intensity regions in the first PHANGS--JWST data and conclude that the level of mid-IR emission appears consistent with arising from atomic, not molecular gas. Also in this issue, \citet{LEROY22JWSTALMAMUSE} compare PHANGS--ALMA CO~(2-1), PHANGS--MUSE H$\alpha$ \citep{PHANGSMUSE22}, and JWST mid-IR at $\sim 70{-}160$~pc scales and make a first attempt at isolating the mid-IR associated with diffuse ISM emission and calibrating the gas-to-mid-IR conversion and separating ISM-tracing emission from emission driven mainly by heating. In a similar effort using a large set of SINGS spectroscopy, \citet{WHITCOMB22MIRCO} examine how different continuum bands and mid-IR emission features correlate with CO and star formation tracers and used this analysis to map out the degree to which parts of the mid-IR spectrum trace either gas or star formation. Taking all of this together, we expect that in the near future, mid-IR emission may offer a new, powerful complement to tradition ISM probes like \ion{H}{1}, CO, or long wavelength dust emission.

\section{Summary and conclusions} \label{sec:summary}

We have measured the observed relations between CO and mid-IR emission based on a large set of integrated and resolved CO and mid-IR measurements for local galaxies. This data set draws heavily on results from a number of surveys, detailed in \S \ref{sec:intdata} and \ref{sec:resdata}. We use these observations to measure the relationships between CO line intensity and mid-IR intensity at $8\mu$m, $12\mu$m, $22\mu$m, and $24\mu$m, as well as a first comparison with JWST results. Table \ref{tab:results} summarizes the results of our correlation analysis. Key points from the analysis are:

\begin{enumerate}
\item There are tight correlations between intensity in all mid-IR bands, $I_{\rm MIR}$, and both CO~(1-0) and CO~(2-1) line-integrated intensity, $I_{\rm CO}$ (\S \ref{sec:scaling}). These correlations hold for integrated galaxies and measurements that break galaxies into radial profiles (median resolution 1.3~kpc; range 0.5{-}1.9~kpc) or individual $17''$ regions (median resolution 1.2~kpc; range 0.3{-}1.8~kpc). 

\item Though the exact ratios and best-fit scaling relations vary by band and line choice (Table \ref{tab:results}), a reasonable zeroth order rule of thumb is that $I_{\rm MIR} \sim 1$~MJy~sr$^{-1}$ when $I_{\rm CO} \sim 1$~K~km~s$^{-1}$ (see \S \ref{sec:normalization}). For standard simple conversions to star formation rate and molecular gas mass this ratio corresponds to a molecular gas depletion time of $\sim 0.8{-}1.7$~Gyr, in excellent agreement with a wide range of recent estimates. We provide more detailed measurements of the median ratio for each band-line combination, as well as the normalization of the best-fit power law relation, in Table \ref{tab:results}, \S \ref{sec:normalization}, and Figure \ref{fig:normalization}. We note that a complicating factor in the normalization of the CO~(1-0) relations is that the sample of local galaxies with single dish CO~(1-0) maps appears biased with respect to the larger sample of galaxies with integrated measurements.

\item Power law scaling relationships offer a good first order description of the relationship between $I_{\rm CO}$ and $I_{\rm MIR}$. We treat the mid-IR intensity as the independent variable because of its high sensitivity compared to CO and derive power laws describing the mean relationship between each band and each line (Table \ref{tab:results}). These can be used to predict CO intensity from mid-IR emission or the reverse. They also capture the empirical relationship that underlies much recent work on the relationship between star formation and molecular gas. We find $I_{\rm CO}$ vs. $I_{\rm MIR}$ slopes in the range $m_{\rm CO-IR} \approx 0.7{-}1.2$ (\S \ref{sec:slope}, Figure \ref{fig:slope}). Inverted to the sense usually adopted to describe star formation scaling relations, $m_{\rm IR-CO}$, we find slopes of $\approx 0.8{-}1.5$, in good agreement with results studying the $\Sigma_{\rm SFR}{-}\Sigma_{\rm mol}$ relation over the last decade. The approximately linear nature of these relations also supports the use of mid-IR emission as a useful empirical predictor of CO emission from galaxies.

\item In general, bands that include a significant PAH feature, $8\mu$m and $12\mu$m, show steeper CO vs.\ mid-IR relations slopes, $m_{\rm CO-IR}$ up to $1.2$, than bands that include only continuum, $22\mu$m or $24\mu$m ( \S \ref{sec:slope}, Figure \ref{fig:slope}). This could reflect that there is a significant contribution of dust mixed with atomic or CO-dark molecular gas to the PAH emission from galaxies, while the $22\mu$m and $24\mu$m may be more directly associated with star formation. Overall the most nearly linear relationship in our analysis is that between CO~(2-1) and $12\mu$m emission, which has $m_{\rm CO-IR} \approx 1.0$ for integrated galaxies and radial profiles and $\approx 0.9$ for individual regions.

\item For any given mid-IR band, the best-fit power laws using CO~(1-0) and CO~(2-1) show moderately different slopes, with CO~(1-0) showing shallower CO vs.\ mid-IR relations slopes, $m_{\rm CO-IR}$, than CO~(2-1) (\S \ref{sec:slope}, Figure \ref{fig:slope}). The typical offset in slope is $\approx 0.2$. Assuming our samples capture a common underlying galaxy population this implies that the CO~(2-1)/CO~(1-0) ratio, $R_{21}$, can be predicted from the large-scale average mid-IR intensity via $R_{21} \propto I_{\rm MIR}^{0.2}$. This is in excellent agreement with recent studies showing $R_{21} \propto \Sigma_{\rm SFR}^{0.15}$ and supports the conclusion that line choice has a corresponding real but modest impact on the slope of derived star formation scaling relations. Assuming that the galaxies with measured CO~(1-0) and CO~(2-1) emission are equivalent, we use these results to construct a statistical predictor of the $R_{21}$ ratio that should be generally useful (Equation \ref{eq:r21}, \S \ref{sec:r21}, Figure \ref{fig:r21}).

\item The ratio of CO to mid-IR emission depends on the integrated properties of a galaxy (\S \ref{sec:intprops}, Figures \ref{fig:intscaling_mass} \& \ref{fig:intscaling_ssfr}, Table \ref{tab:results}). $I_{\rm CO}/I_{\rm MIR}$ weakly correlates with stellar mass ($M_\star$) and anti-correlates with specific star formation rate (SFR/$M_\star$). The stellar mass trend agrees with results showing higher SFR-per-CO in low mass galaxies and has the sense expected if more mid-IR emission emerges from atomic or CO-dark molecular gas in low mass, low metallicity galaxies. The specific star formation rate trend is consistent with recent results that link starburst and quenching to the rate of star formation per unit molecular gas (rather than only the gas supply).

\item We compare the first mid-IR mapping results from PHANGS--JWST to CO~(2-1) mapping from PHANGS--ALMA to make an initial placement of JWST data into these relations (Table \ref{tab:jwst}, Figures \ref{fig:phangsjwst_sample}, and \ref{fig:phangsjwst_data}). As a sample, the PHANGS--JWST targets do a reasonable job of reflecting the larger population of previously mapped galaxies (Figure \ref{fig:phangsjwst_sample}. Bearing in mind that our initial JWST data cover only four galaxies and spans a limited range of intensities, we show good agreement between the slopes relating CO~(2-1) to F770W, F1000W, or F1130W in the first four PHANGS--JWST galaxies and that relating CO~(2-1) to $8\mu$m in our larger data set (Table \ref{tab:jwst}). Similarly, the CO~(2-1) to F2100W relationship seen combining PHANGS--ALMA with JWST resembles the CO~(2-1) to $24\mu$m relationship seen for the larger sample. The slope relating CO~(2-1) the PAH-tracing bands (F770W, F1130W, and likely F1000W) is $\sim 1.2{-}1.4$, implying more mid-IR relative to CO at low intensities. This offers some support for the idea that the mid-IR emission, especially the PAH emission, emerges from a mixture of phases that include non CO-emitting gas and that mid-IR JWST observations have potential applications to trace multiple phases of the ISM.

\end{enumerate}

Finally, we remark that given the heterogeneity of our data and the breadth of targets studied, the overall strength of the correlation between mid-IR and CO emission is notable. Though not perfectly linear, the correlation appears to be among the strongest in extragalactic astronomy. This likely reflects that both mid-IR and CO emission act partially as gas tracers and partially as star formation rate tracers to a greater extent than generally appreciated. Our goal is that this study helps frame the detailed exploration of mid-infrared emission from galaxies enabled by the revolutionary sensitivity and resolution of the JWST and the powerful synergy of JWST and ALMA. 

\section*{Acknowledgments}

We thank the anonymous referee for a timely and constructive report that improved the paper.

This work builds on the hard work of a number of scientific teams and we gratefully acknowledge all of the teams with data sets cited in Sections \ref{sec:intro} and \ref{sec:data} for making their data and products public. The work was specifically carried out in the context of the PHANGS and EDGE collaborations.

AKL gratefully acknowledges support by grants 1653300 and 2205628 from the National Science Foundation, by award JWST-GO-02107.009-A, and by a Humboldt Research Award from the Alexander von Humboldt Foundation.

ADB acknowledges support by NSF-AST2108140 and award JWST-GO-02107.008-A.

KS acknowledges support by JWST-GO-02107.006-A and National Science Foundation grant 2108081.

ER acknowledges the support of the Natural Sciences and Engineering Research Council of Canada (NSERC), funding reference number RGPIN-2022-03499.

JMDK gratefully acknowledges funding from the European Research Council (ERC) under the European Union's Horizon 2020 research and innovation programme via the ERC Starting Grant MUSTANG (grant agreement number 714907). COOL Research DAO is a Decentralized Autonomous Organization supporting research in astrophysics aimed at uncovering our cosmic origins.

MC gratefully acknowledges funding from the DFG through an Emmy Noether Research Group (grant number CH2137/1-1).

JK gratefully acknowledges funding from the Deutsche Forschungsgemeinschaft (DFG, German Research Foundation) through the DFG Sachbeihilfe (grant number KR4801/2-1).

EJW acknowledges the funding provided by the Deutsche Forschungsgemeinschaft (DFG, German Research Foundation) -- Project-ID 138713538 -- SFB 881 (``The Milky Way System'', subproject P1). 

FB would like to acknowledge funding from the European Research Council (ERC) under the European Union’s Horizon 2020 research and innovation programme (grant agreement No.726384/Empire)

MB acknowledges support from FONDECYT regular grant 1211000 and by the ANID BASAL project FB210003.

RSK acknowledges funding from the European Research Council via the ERC Synergy Grant ``ECOGAL'' (project ID 855130), from the Deutsche Forschungsgemeinschaft (DFG) via the Collaborative Research Center ``The Milky Way System''  (SFB 881 -- funding ID 138713538 -- subprojects A1, B1, B2 and B8) and from the Heidelberg Cluster of Excellence (EXC 2181 - 390900948) ``STRUCTURES'', funded by the German Excellence Strategy. RSK also thanks the German Ministry for Economic Affairs and Climate Action for funding  project ``MAINN'' (funding ID 50OO2206). 

KK gratefully acknowledges funding from the Deutsche Forschungsgemeinschaft (DFG, German Research Foundation) in the form of an Emmy Noether Research Group (grant number KR4598/2-1, PI Kreckel).

DC acknowledges support by the German \emph{Deut\-sche For\-schungs\-ge\-mein\-schaft, DFG\/} project number SFB956A.

JC acknowledges support from ERC starting grant \#851622 DustOrigin.

RCL acknowledges support provided by a NSF Astronomy and Astrophysics Postdoctoral Fellowship under award AST-2102625.

EWK acknowledges support from the Smithsonian Institution as a Submillimeter Array (SMA) Fellow and the Natural Sciences and Engineering Research Council of Canada (NSERC).

MQ acknowledges support from the Spanish grant PID2019-106027GA-C44, funded by MCIN/AEI/10.13039/501100011033.

JS acknowledges support from NSERC through a Canadian Institute for Theoretical Astrophysics (CITA) National Fellowship. The research of CDW is supported by grants from NSERC and the Canada Research Chairs program.

M.R. wishes to acknowledge support from ANID(CHILE) through FONDECYT grant No1190684 and partial support from ANID Basal FB210003

ES and TGW acknowledge funding from the European Research Council (ERC) under the European Union’s Horizon 2020 research and innovation programme (grant agreement No. 694343).

JPe acknowledges support by the DAOISM grant ANR-21-CE31-0010 and by the Programme National ``Physique et Chimie du Milieu Interstellaire'' (PCMI) of CNRS/INSU with INC/INP, co-funded by CEA and CNES.

This work uses the $z=0$ Multiwavelength Galaxy Synthesis, DOI 10.26131/IRSA6 \citep{z0mgsdoi}.

This work is based in part on observations made with the NASA/ESA/CSA James Webb Space Telescope. The data were obtained from the Mikulski Archive for Space Telescopes at the Space Telescope Science Institute, which is operated by the Association of Universities for Research in Astronomy, Inc., under NASA contract NAS 5-03127 for JWST. These observations are associated with program 2017. The specific observations analyzed can be accessed via \dataset[ 10.17909/9bdf-jn24]{http://dx.doi.org/10.17909/9bdf-jn24}.

This paper makes use of the following ALMA data: \\
\noindent ADS/JAO.ALMA\#2012.1.00650.S, \linebreak % (N628/M74)
ADS/JAO.ALMA\#2013.1.00803.S, \linebreak % (N5128/CenA)
ADS/JAO.ALMA\#2013.1.01161.S, \linebreak % (N1365 + N5236/M83)
ADS/JAO.ALMA\#2015.1.00121.S, \linebreak % (N5236/M83)
ADS/JAO.ALMA\#2015.1.00782.S, \linebreak % (N1313 + N7793)
ADS/JAO.ALMA\#2015.1.00925.S, \linebreak % (pilot low mass)
ADS/JAO.ALMA\#2015.1.00956.S, \linebreak % (pilot high mass)
ADS/JAO.ALMA\#2016.1.00386.S, \linebreak % (N5236/M83)
ADS/JAO.ALMA\#2017.1.00392.S, \linebreak % (low mass follow-up)
ADS/JAO.ALMA\#2017.1.00766.S, \linebreak % (early-type)
ADS/JAO.ALMA\#2017.1.00886.L, \linebreak % (large program)
ADS/JAO.ALMA\#2018.1.01321.S, \linebreak % (N253, N300, Circinus)
ADS/JAO.ALMA\#2018.1.01651.S, \linebreak % (main sample follow-up)
ADS/JAO.ALMA\#2018.A.00062.S, \linebreak % (ACA-only nearby)
ADS/JAO.ALMA\#2019.1.01235.S, \linebreak % (local sample follow up)
ADS/JAO.ALMA\#2019.1.00763.L, \linebreak % (VERTICO)
ADS/JAO.ALMA\#2019.2.00129.S, \linebreak % (N1068)
ALMA is a partnership of ESO (representing its member states), NSF (USA), and NINS (Japan), together with NRC (Canada), NSC and ASIAA (Taiwan), and KASI (Republic of Korea), in cooperation with the Republic of Chile. The Joint ALMA Observatory is operated by ESO, AUI/NRAO, and NAOJ. The National Radio Astronomy Observatory is a facility of the National Science Foundation operated under cooperative agreement by Associated Universities, Inc.

\vspace{5mm}
\facilities{ALMA, APEX, IRAM:30m, FCRAO, CARMA, ARO:12m, ARO:SMT, No:45m, BIMA, WISE, Spitzer (IRAC, MIPS), JWST (MIRI)}

\software{astropy \citep{ASTROPY13,ASTROPY18,ASTROPY22}}

% Note - adjusted for arxiv submission. Switch back for general case.
%\bibliography{temp}{}
\bibliography{main.bbl}
\bibliographystyle{aasjournal}

\suppressAffiliationsfalse
\allauthors

\end{document}